\documentclass[a4paper,fleqn,usenatbib]{mnras}

\usepackage{newtxtext,newtxmath}

\usepackage[T1]{fontenc}
\usepackage{ae,aecompl}

\usepackage{graphicx}	
\usepackage{amsmath}	
\usepackage{amssymb}	

\usepackage[dvipsnames]{xcolor}

\title[Searching for HI imprints in cosmic web filaments]{Searching for HI imprints in cosmic web filaments with 21-cm intensity mapping}

\author[D. Tramonte et al.]{
Denis Tramonte,$^{1,2}$\thanks{E-mail: TramonteD@ukzn.ac.za}
Yin-Zhe Ma,$^{1,2}$
Yi-Chao Li,$^{1,3}$
Lister Staveley-Smith$^{4,5}$
\\ 
\\
$^{1}$School of Chemistry and Physics, University of KwaZulu-Natal, Westville Campus, Private Bag X54001, Durban, South Africa \\
$^{2}$NAOC-UKZN Computational Astrophysics Center (NUCAC), University of Kwazulu-Natal, Durban, 4000, South Africa \\
$^{3}$Department of Physics \& Astronomy, University of the Western Cape, Cape Town 7535, South Africa \\
$^{4}$International Centre for Radio Astronomy Research (ICRAR), M468, The University of Western Australia, 35 Stirling Highway, Crawley, WA 6009, Australia \\
$^{5}$ARC Centre of Excellence for All Sky Astrophysics in 3 Dimensions (ASTRO 3D)
}

\date{Accepted 2019 July 31. Received 2019 July 30; in original form 2019 May 25}

\pubyear{2019}

\begin{document}
\label{firstpage}
\pagerange{\pageref{firstpage}--\pageref{lastpage}}
\maketitle

\begin{abstract}
	We investigate the possible presence of neutral hydrogen (HI) in intergalactic filaments at very low redshift ($z\sim 0.08$), by stacking a set of 274,712 2dFGRS galaxy pairs over 21-cm maps obtained with dedicated observations conducted with the Parkes radio telescope, over a total sky area of approximately 1,300 square degrees covering two patches in the northern and in the southern Galactic hemispheres. The stacking is performed by combining local maps in which each pair is brought to a common reference frame; the resulting signal from the edge galaxies is then removed to extract the filament residual emission. We repeat the analysis on maps cleaned removing either 10 or 20 foreground modes in a principal component analysis. 
	Our study does not reveal any clear HI excess in the considered filaments in either case; we determine upper limits on the total filament HI brightness temperature at $T_{\rm b} \lesssim 10.3 \,\mu\text{K}$ for the 10-mode and at $T_{\rm b} \lesssim 4.8 \,\mu\text{K}$ for the 20-mode removed maps at the 95\% confidence level. These estimates translate into upper limits for the local filament HI density parameter, $\Omega_{\rm HI}^{\rm (f)} \lesssim 7.0\times10^{-5}$ and $\Omega_{\rm HI}^{\rm (f)} \lesssim 3.2\times10^{-5}$ respectively, and for the HI column density, $N_{\rm HI} \lesssim 4.6\times10^{15}\,\text{cm}^{-2}$ and $N_{\rm HI} \lesssim 2.1\times10^{15}\,\text{cm}^{-2}$ respectively. 
	These column density constraints are consistent with previous detections of HI in the warm-hot intergalactic medium obtained observing broad Ly $\alpha$ 
	absorption systems. The present work shows for the first time how such constraints can be achieved using the stacking of galaxy pairs on 21-cm maps. 
\end{abstract}

\begin{keywords}
cosmology: large-scale structure of Universe -- radio lines: ISM -- ISM: general
\end{keywords}



\defcitealias{anderson18}{A18}
\defcitealias{tanimura19}{T19}
\defcitealias{degraaff19}{DG19}

\begin{table*}
\centering
	\caption{Information on the six Parkes patches used in our analysis. For right ascension (RA), declination (Dec) and frequency ($\nu$) we report the patch central value and the resolution expressed in terms of the pixel size. The RA pixel size is quoted in degrees of polar rotation, thus being larger for the patches with low declination. All patches have the same number of pixels, namely 486 pixels in RA, 106 pixels in Dec and 64 pixels in $\nu$. The last columns report the number of 2dFGRS galaxies and pairs used for each patch.}
\label{tab:parkes}	
\setlength{\tabcolsep}{1.2em}
\begin{tabular}{p{0.7cm}p{0.02cm}p{1cm}p{1cm}p{0.02cm}p{1cm}p{1cm}p{0.02cm}p{1cm}p{1cm}p{0.02cm}p{1cm}p{1cm}}
\hline
\hline
Patch & & \multicolumn{2}{c}{RA [deg]} & & \multicolumn{2}{c}{Dec [deg]} & & \multicolumn{2}{c}{$\nu$ [MHz]} & & Galaxy & Pair \\ 
\\[-1em]
\cline{3-4}\cline{6-7}\cline{9-10}
\\[-1em]
 & & Centre & Resolution & & Centre & Resolution & & Centre & Resolution & & number & number \\
\hline
1 & & -18.0 & 0.092 & & -29.75 & 0.080 & & 1315.5 & 1.0 & & 2209 & 39214 \\ 
2 & & 33.0 & 0.092 & & -29.75 & 0.080 & & 1315.5 & 1.0 & & 2438 & 40153 \\ 
3 & & 165.0 & 0.080 & & -0.05 & 0.080 & & 1315.5 & 1.0 & & 2658 & 52950 \\ 
4 & & 182.0 & 0.080 & & -0.05 & 0.080 & & 1315.5 & 1.0 & & 2688 & 57649 \\ 
5 & & 199.0 & 0.080 & & -0.05 & 0.080 & & 1315.5 & 1.0 & & 2542 & 57638 \\ 
6 & & 216.0 & 0.080 & & -0.05 & 0.080 & & 1315.5 & 1.0 & & 1444 & 27108 \\ 
\hline
\hline
\end{tabular}
\end{table*}

\section{Introduction}
\label{sec:introduction}

According to the flat $\Lambda$-cold-dark matter ($\Lambda$CDM) cosmological model, baryonic matter 
is expected to account for approximately 4.9\% of the total energy density in the 
Universe, as confirmed by both the observation of cosmic microwave background (CMB) 
anisotropies~\citep{hinshaw13, planck18} and, with lower accuracy, 
by the models of Big-Bang nucleosynthesis based on the observed abundance of 
light elements~\citep{fields14, cooke18_1, cooke18_2}. Direct observations confirm these prediction 
at relatively high redshifts ($z \gtrsim 2$), where most of the baryons 
 are detected via the Lyman-$\alpha$ absorption 
of radiation from distant quasars~\citep{rauch97, weinberg97, rauch98}. At lower 
redshift, however, the joint contribution of baryons in both condensed phase (stars) and diffuse phase (interstellar medium, intracluster medium) does not match the 
predicted abundance, accounting only for roughly 50\% of the expected amount of 
baryonic matter~\citep{persic92, fukugita98, fukugita04, bregman07, nicastro08, shull12}. This 
discrepancy has traditionally been referred to as the ``missing baryon problem'', 
and its understanding is crucial in proving the validity of the standard 
cosmological model. 

Results from hydrodynamical simulations suggest that these missing baryons are 
to be found in a non-virialised diffuse phase aligned with the filaments in the 
cosmic web, gravitationally heated during the process of structure 
formation~\citep{cen99, dave01, cen06}. This warm-hot intergalactic medium (WHIM) 
is expected to have temperature ranging in between $10^5$ and $10^7\,\text{K}$ and density around 
ten times the mean baryon density. In the past two decades many searches have been 
dedicated to the direct detection of the missing baryons. The hottest WHIM 
component can be probed with X-ray observations, as it was done in~\citet{kull99}, 
\citet{scharf00}, \citet{zappacosta02}, \citet{werner08}, \citet{ibaraki14} 
and~\citet{eckert15}. These studies focused on a particular 
structure or pointing of the corresponding X-ray survey. Other studies have conducted searches for 
the imprint of missing baryons using CMB data, utilising in particular the 
Sunyaev-Zel'dovich (SZ) effect produced by the 
WHIM~\citep{ursino14, van_waerbeke14, genova_santos15, ma15, hernandez_monteagudo15}. 
A novel approach to observe the WHIM SZ emission was considered 
by~\citet{tanimura19} and~\citet{degraaff19} (hereafter T19 and DG19, respectively). 
Because of the diffuse nature of the WHIM, the SZ imprint from an individual 
filament would be too faint to be detected; these authors therefore reverted to 
stacking the contribution from several large-scale structure (LSS) filaments found in between 
suitable pairs of SDSS-DR12 galaxies~\citep{adam15} over the Planck Compton 
parameter ($y$) maps~\citep{planck_sz15}, analysing this way a large sky area 
($\sim 13,600$ square degrees). The two studies agree in the level of the 
filament Compton parameter excess and in the detection significance, although 
probing different redshift ranges, and clearly prove observationally the presence 
of hot baryons in the large-scale filaments. 

Because of the low surface brightness expected from the WHIM diffuse emission, 
other work has tackled the search for absorption features in UV and X-ray 
wavelengths produced by highly ionised metals, among which oxygen is the most 
relevant given its relatively high cosmic abundance. Assumptions on the local 
metallicity allow estimation in this way of the WHIM baryon 
content~\citep{fang07, buote09, zappacosta10, nicastro13}. A similar approach 
is the search for absorption features from neutral hydrogen (HI) in WHIM, using 
observations in 
the far-UV~\citep{richter06, danforth10, pessa18}. Although the WHIM is mostly ionised, 
a small fraction of the gas, in the range 
$10^{-6} \lesssim x_{\text{HI}} \lesssim 10^{-5}$, is expected to be 
neutral~\citep{sutherland93}. This HI component can be detected via 
Lyman-$\alpha$ absorption; due to the WHIM high temperature, the line is 
expected to be very broadened, so these neutral components are usually referred 
to as broad Lyman-$\alpha$ (BLA) absorbers. Under the assumption of collisional 
ionised equilibrium (CIE) the line width allows the estimation of the total 
hydrogen column density and hence of the local baryonic content. Although this 
method is mostly sensitive to the regions of the WHIM with relatively low temperature 
and high gas column density, and the detection is subject to a number of 
uncertainties (mainly deviations from CIE and non-thermal line broadening 
processes), \citet{richter06} provided the value 
$\Omega_{\rm b}(\text{BLA})\simeq 0.0027\,h_{70}^{-1}$ as a reliable lower 
limit for the BLA baryonic content, implying that a significant fraction of 
baryons in the local Universe are traced by these residual neutral 
components. 
\begin{figure*}
\includegraphics[trim= 10mm 0mm 0mm 0mm, scale=0.33]{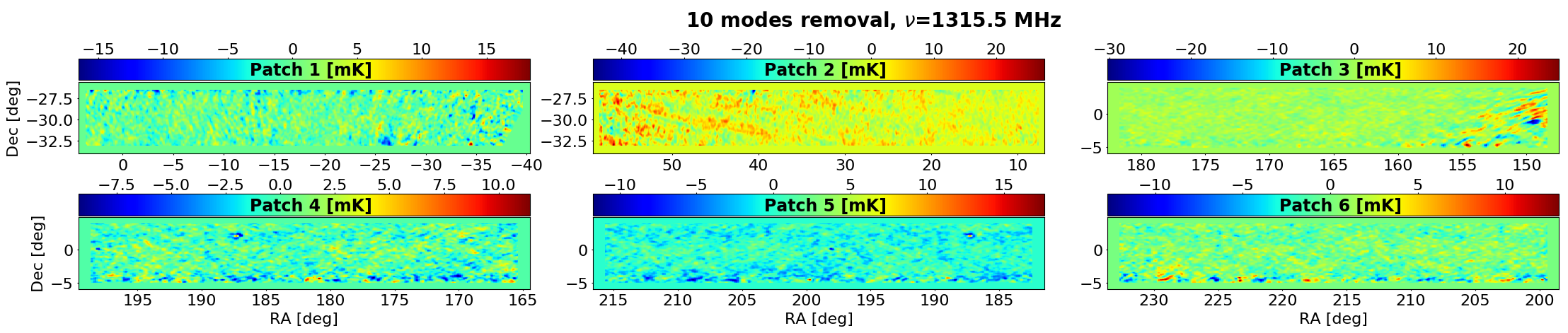}
\newline
\newline
\includegraphics[trim= 10mm 0mm 0mm 0mm, scale=0.33]{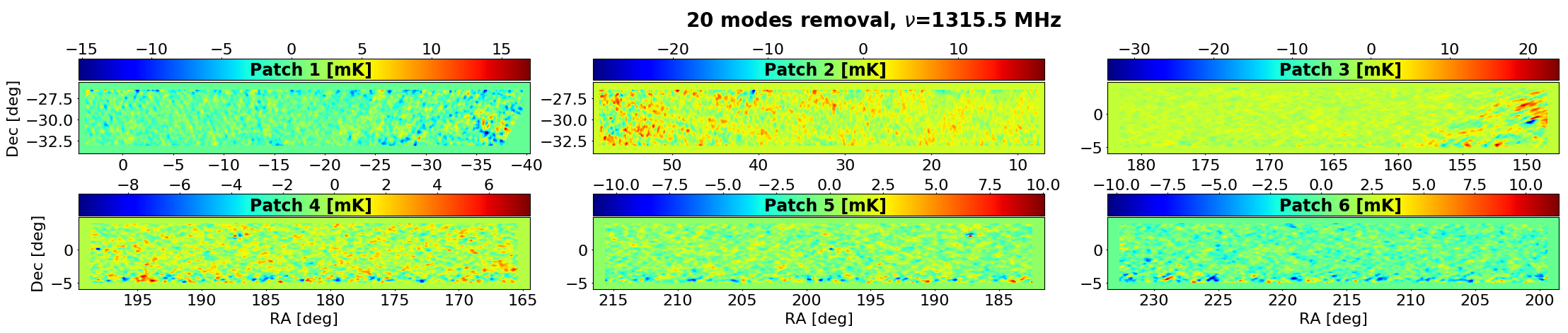}
\newline
\newline
\includegraphics[trim= 10mm 15mm 0mm 0mm, scale=0.33]{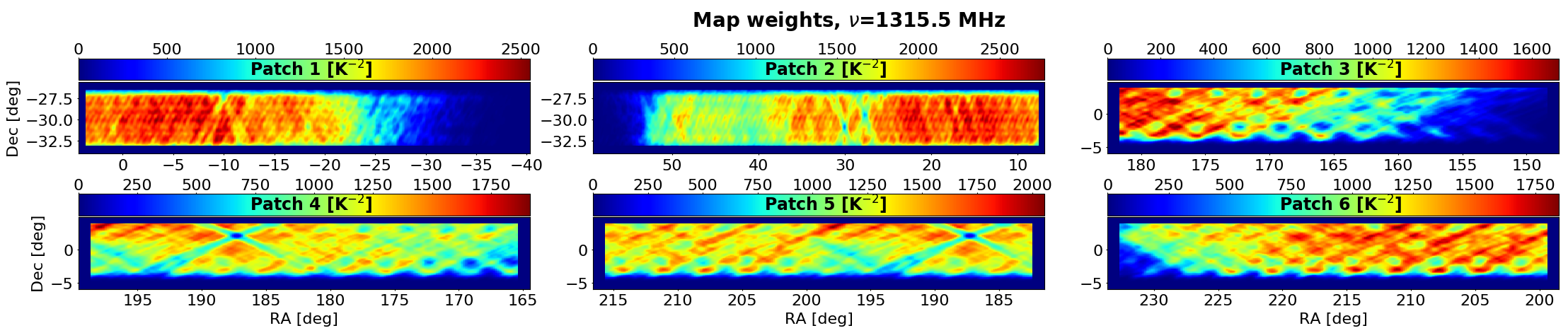}
	\caption{An example of the Parkes data used in this work. 
	\textit{Top panel}. 10-mode foreground removed 21-cm intensity maps shown 
	in units of mK for all six patches in the (RA, Dec) frame for a single 
	frequency/redshift slice, corresponding to the centre of the Parkes 
	Multibeam Receiver bandwidth. \textit{Middle panel}. Same as 
	above, but showing the 20-mode foreground removed maps. 
	\textit{Bottom panel}. Plot of the map weights per patch in units 
	$\text{K}^{-2}$ for the same frequency slice shown in the upper panels. 
	Features clearly visible in these plots include the masking of local bright 
	radio sources and the crossing of differently oriented stripes 
	corresponding to the raster scanning of the telescope at 
	constant elevation. }
\label{fig:parkes}	
\end{figure*}

The present work aims at contributing to the search for neutral hydrogen in the 
local Universe, focussing this time on the 21-cm line from the spin-flip 
hyperfine transition in the hydrogen ground state~\citep{pritchard12}. Compared 
to the Lyman-$\alpha$ line, which for low redshift sources falls in the UV 
range and is therefore blocked by the atmosphere, the 21-cm emission lies in 
the microwave range which enables its detection using ground-based 
facilities.~\citet{pen09} first showed the feasibility of tracing cosmic 
structures using 21-cm maps. Subsequent works relied on the cross-correlation 
between 21-cm and galaxy maps spectra to obtain estimates of the HI energy 
density parameter~\citep{chang10, masui13}, agreeing in the estimate 
$\Omega_{\rm HI} \sim 6\times10^{-4}$ at $z \sim 0.8$. The correlation between 
galaxies and HI at very low redshfit was also investigated 
in~\citet[hereafter A18]{anderson18}. 
This approach for estimating the HI mass density relies on the 
expected full correlation between HI fluctutations and galaxy overdensities on large scales.
The present study, conversely, targets specifically the cosmic web filaments, 
and aims at searching for the 21-cm 
signature from the neutral gas fraction in the WHIM. The feasibility of 
a direct detection of 21-cm emission from the filamentary gas has been 
explored in~\citet{kooistra17}.
In this work, however, we adopt a 
methodology very similar to the one used by~\citetalias{tanimura19} 
and~\citetalias{degraaff19}, based on the stacking 
of galaxy pairs to enhance the signal coming from the large-scale filaments, 
and we employ the same data set used by~\citetalias{anderson18}, namely the 21-cm maps from the 
Parkes telescope and the galaxy catalogue from the Two-Degree-Field (2dF) survey. 
The final goal is to assess the amount of HI found in filaments at very low redshift ($z<0.1$), 
by providing estimates (or upper limits) on the HI column density $N_{\rm HI}$ and on the locally defined HI 
density parameter $\Omega_{\text{HI}}^{\rm (f)}$ (the superscript ``f'' is used to stress that this quantity 
is evaluated in filaments, and to distinguish it from the global HI density parameter in the Universe $\Omega_{\text{HI}}$).
Such a result is complementary to the estimates based on the detection of the 
21-cm emission proceeding from the HI confined inside galaxies, and shows the 
relative amount of neutral baryons that are to be found as a diffuse component 
in the cosmic web. 

The paper is organised as follows. In Section~\ref{sec:data} we describe the 
data set we employed and the pre-processing that was required by 
the subsequent analysis. The procedure for the extraction of the filamentary 21-cm 
signal and the determination of its uncertainty is detailed in 
Section~\ref{sec:methodology}. In Section~\ref{sec:discussion_stack} we discuss the 
significance of the results obtained from the galaxy-pair stacking.
Section~\ref{sec:hi} is dedicated to the relation 
between the derived filament temperature and the corresponding local HI content. 
Finally, Section~\ref{sec:conclusions} presents the conclusions. Whenever the 
computation of cosmological quantities is required, throughout this work we 
adopt a flat $\Lambda$CDM cosmology with $\Omega_{\rm m}=0.3$, $\Omega_{\rm b}=0.049$, 
$\Omega_{\Lambda}=0.7$ and $H_0=70\,\text{km}\,\text{s}^{-1}\text{Mpc}^{-1}$.

\section{Data set}
\label{sec:data}
We describe in this section the data set employed for our analysis. In order to 
detect the HI emission from intergalactic filaments we need a galaxy catalogue to 
identify the position and orientation of the filaments, and a sufficiently extended 
sky map in 21-cm to extract the corresponding signal. 
Clearly, the map and the catalogue must have a significant overlap in both angular 
and redshift coverage. We concluded that, in terms of the total covered sky area, 
the best choice that suits our purpose is a set of 21-cm maps obtained 
with the Parkes telescope with tailored observations that covered the region 
spanned by the 2dF Galaxy Catalogue. 
We describe both the maps and the catalogue in details in the following. 

\subsection{21-cm maps}
\label{ssec:parkes}
We employ the 21-cm intensity maps from the Parkes Observatory described in~\citetalias{anderson18}. 
The starting point for our work are 
the foreground-cleaned maps, 
which can be considered to be representative of the HI emission alone.
We redirect to the aforementioned reference for a detailed description of the full 
data processing involved in the production of the maps, including 
observational strategy, bandpass calibration, mapmaking algorithm and foreground 
removal. We limit this section to a description of the final product which is 
relevant for the subsequent analysis.
\begin{figure*}
\includegraphics[trim= 20mm 0mm 0mm 0mm, scale=0.2]{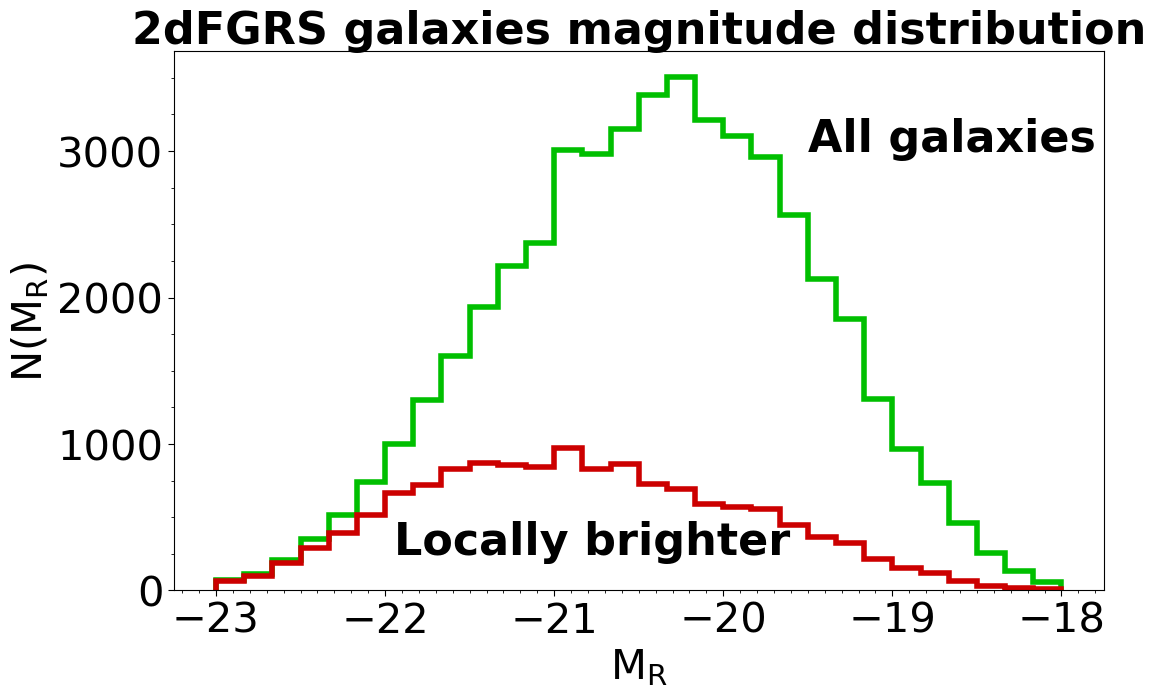}
\includegraphics[trim= 0mm 0mm 0mm 0mm, scale=0.2]{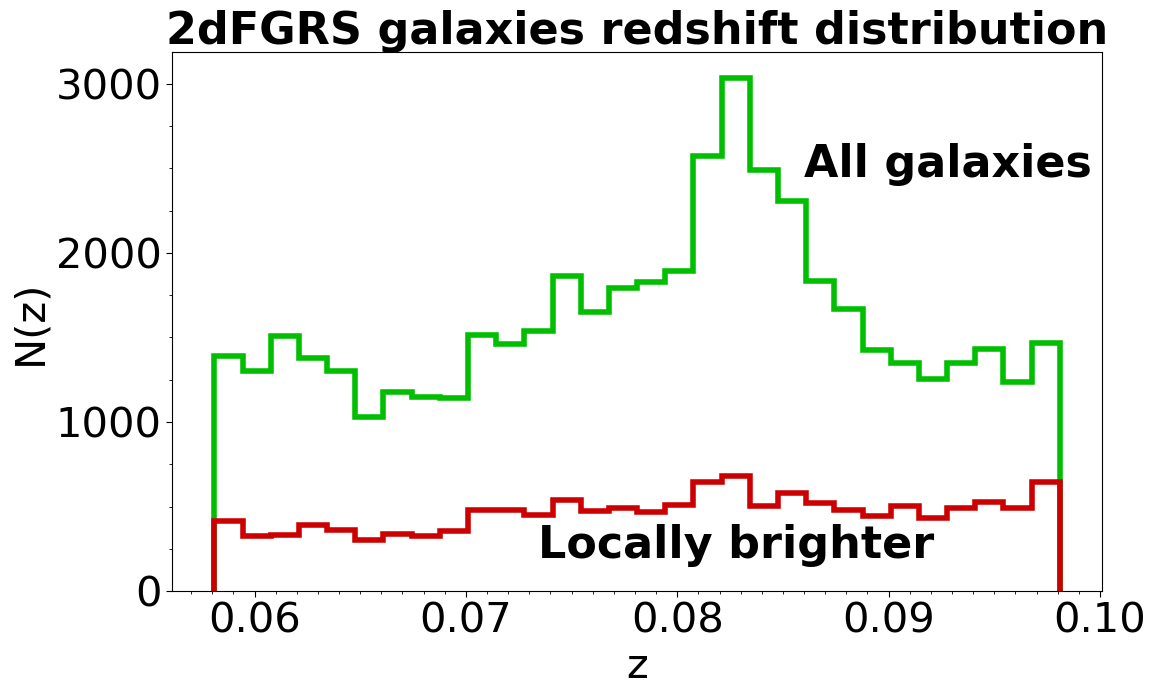}
\includegraphics[trim= 0mm 0mm 0mm 0mm, scale=0.2]{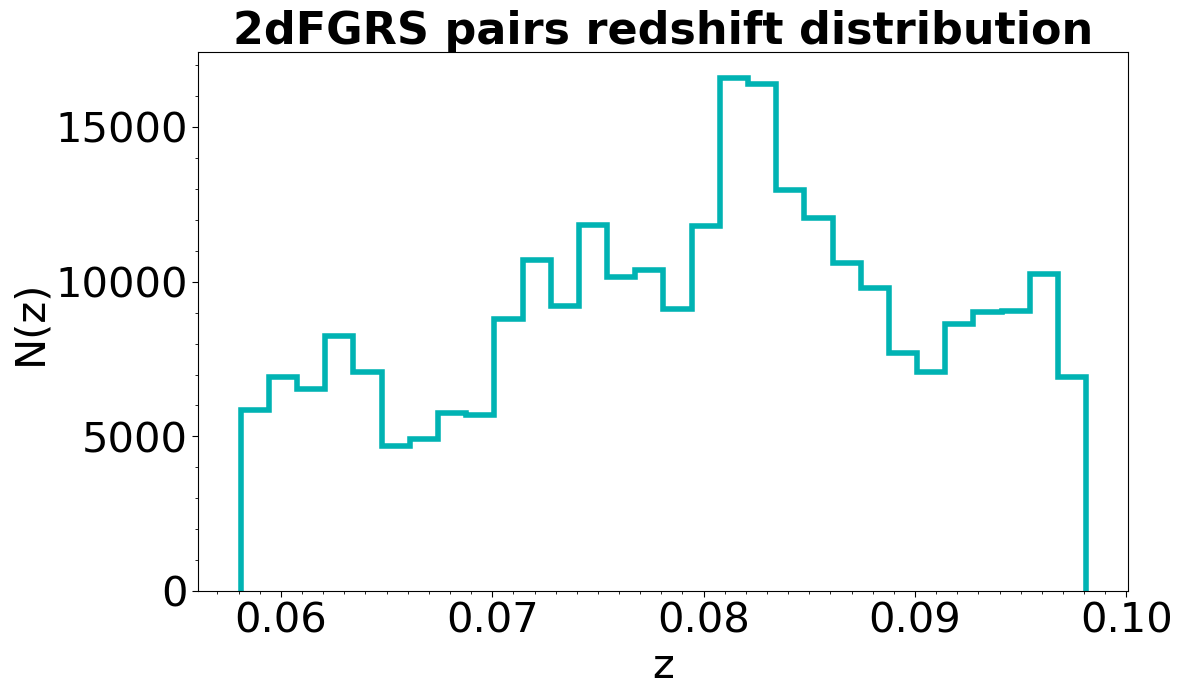}
	\caption{\textit{Left panel.} Effect of the selection of locally brighter galaxies 
	on the original 2dFGRS catalogue included in the Parkes data volume, in terms of 
	the catalogue distribution in absolute R-band magnitude. \textit{Central 
	panel}. Same as in the left panel but showing the galaxy redshift distribution. 
	\textit{Right panel.} Redshift distribution for the final pair catalogue 
	used in this work.}
\label{fig:2df}	
\end{figure*}

Data employed for the map-making came from a set of observations taken using 
the Parkes 21-cm Multibeam Receiver~\citep{staveley_smith96} during 
a single week between April and May 2014, for a total observing time of 152 hours. 
The full Multibeam Correlator was employed with 1024 frequency channels 
over a full bandwith of 64 MHz centred at 1315.5 GHz, resulting in a 62.5 kHz frequency resolution.
The Parkes beam sets the angular resolution at 14 arcmin.
These observations aimed at covering the area spanned by the 2dF Galaxy Catalogue: 
the resulting maps cover two stripes, one in the northern  
and one in the southern Galactic hemisphere, resulting in a total sky coverage 
of $\sim 1,300$ square degrees. In order to improve the computational 
efficiency of the mapmaking code, these two main fields had been further divided 
into a total of six sub-fields, two in the southern and four 
in the northern Galactic hemisphere. Hereafter we shall label as 
``1'' and ``2'' the southern patches, and from ``3'' to ``6'' the northern patches
(see Fig.~\ref{fig:parkes}). 
Each patch was delivered in the form of a standard three-dimensional matrix: the 
three axes correspond to the J2000.0 equatorial coordinates and to frequency, 
(RA, Dec, $\nu$). The angular pixel size is 0.08 deg, while the frequency resolution  
 was degraded to 1 MHz after local radio frequency interference removal. The resulting sampling 
 in terms of pixel numbers is (486, 106, 64), which is common to all patches.  
Information about the spatial and frequency coverage for 
each patch, together with the corresponding pixel resolution, can be found in 
Table~\ref{tab:parkes} (notice that the frequency range
was the same for all the patches, being determined by the receiver configuration 
chosen for this set of observations). 
 
At these frequencies the radio foregrounds, mainly Galactic and extra-Galactic 
synchrotron emission, are two to three orders of magnitude brighter than the 
21-cm emission. As mentioned, these maps are already foreground-cleaned; 
the identification and removal of these foregrounds at the map 
level is described in~\citetalias{anderson18} and is based on the algorithm detailed 
in~\citet{switzer15}. Given the dominant amplitude of the foregrounds, and their
expected frequency smoothness compared to the 21-cm signal~\citep{liu12},
the cleaning is based on a Principal Component Analysis (PCA) in which the 
higher frequency correlated modes are removed. The result will depend on the 
choice of the number of modes to be removed, taking into account that although each 
of these modes is dominated by foregrounds, it also carries a contribution from the 
21-cm signal; consequently, the removal of a high number of modes yields 
cleaner maps but also implies a higher loss of HI signal. In~\citetalias{anderson18} the 10-mode 
removed maps are used, and we will adopt the same choice for the present work. 
However, in order to assess the effect of the local residual foregrounds on 
our final estimates, we will also conduct the same analysis independently on 
the 20-mode removed maps. 

In the top and middle panels of Fig.~\ref{fig:parkes}
we show the six patches 21-cm maps sliced at the central frequency of the 
Multibeam Receiver, for the cases of 10 and 20 foreground removal modes. 
The maps show both positive and negative structures at 
the $10\,\text{mK}$ level. It is clear from the right-ascension ranges that the 
northern patches are partially overlapping; however, each patch undergone 
independent foreground removal, and as a result the structure pattern is not 
the same comparing two different patches in their overlapping region, the 
difference being again of the order of a few mK. For this reason, 
in this work we will analyse each patch independently, which will allow us to
establish the consistency of our results across the different Parkes fields.
Additionally, we will present the results obtained by combining all the 
patches together, as the case with the highest statistical significance. 

Apart from these 21-cm maps we were also provided with an analogous set of weight 
maps, which account for the non-homogeneous coverage of the observed fields. The 
weight maps are also split into six patches and are issued in the same resolution 
and pixelisation as the signal maps. The value stored in each pixel is the 
corresponding inverse squared noise weight, which is roughly proportional to 
the time spent observing it. The bottom panel of 
Fig.~\ref{fig:parkes} shows the weight maps for a single frequency slice; 
the plot allows to recognise different structures, like the masking of local 
bright radio sources, appearing as dark spots, and the observation strategy, 
which produces bright stripes with different orientations. Indeed, 
data were acquired using azimuth scans at constant elevation, 
and the stripe orientation depends on the apparent motion of the field 
(rising or setting) during each observation.  
The highly spatial inhomogeneous amplitude of 
the weight maps is to be taken into account when combining the 21-cm intensities 
extracted from different regions of the signal maps.

\subsection{Galaxy and pair catalogues}
\label{ssec:2dfgalaxies}
The Two-Degree-Field Galaxy Redshift Survey~\citep[2dFGRS,][]{colless01} is a 
spectroscopic survey conducted between 1997 and 2002 using the 3.9 meter telescope 
at the Anglo-Australian Observatory; its data were released in June 2003. The 
survey provides spectroscopic redshifts for 245,591 sources distributed over 
two fields, one in the northern and one in the southern Galactic hemisphere, 
plus a set of additional random pointings scattered around the southern field. 
We downloaded the 2dFGRS ``Catalogue of best spectroscopic observations''
from the official website\footnote{\url{http://www.2dfgrs.net/}} and performed a 
set of queries to extract the sources most suitable to our analysis.  

We perform a first selection by considering only the sources with a reliable 
redshift determination, quantified by a quality parameter $q$ ranging from 1 
to 5 (according to the 2dF documentation, only the sources with $q \geq 3$ 
can be considered reliable); we also discard the sources with a negative 
redshift estimation. This lowers the total number of sources to 227,190. 
Secondly, the catalogue is queried to select only the galaxies located within 
the angular and redshift span of the Parkes patches described in Section~\ref{ssec:parkes}. 
In order to avoid possible strong residual contaminations from the patch edges, 
we also discard sources located within a Dec limit of $1\,\text{deg}$ and a RA limit 
of $3\,\text{deg}$ from each patch boundaries (the difference between these two values 
comes from the patch extensions in RA being a factor $\sim 3$ larger than the extensions 
in Dec). 
We also discard galaxies with redshift within the lowest 10 MHz 
and the highest 4 MHz of the Parkes frequency range, which according to~\citetalias{anderson18} 
are also affected by higher noise due to band-edge effects. At this point the initial 2dFGRS 
catalogue is split into a set of six sub-catalogues, one for each Parkes patch. For the case 
of the overlapping Parkes fields, galaxies belonging to two patches are included in both the 
corresponding sub-catalogues. Indeed, as already mentioned, the analysis for each patch is conducted independently, 
and this choice improves the available statistics; when combining estimations from different patches, 
the resultant multiplicity is accounted for by using a proper weighting, as it is detailed in 
Section~\ref{ssec:stacking}. In total, the restriction to the patch areas reduces the catalogue to a total 
of 36,800 galaxies, which amounts to 48,340 objects if we count each galaxy with its multiplicity.
\begin{figure*}
\includegraphics[trim= 20mm 0mm 10mm 0mm, scale=0.18]{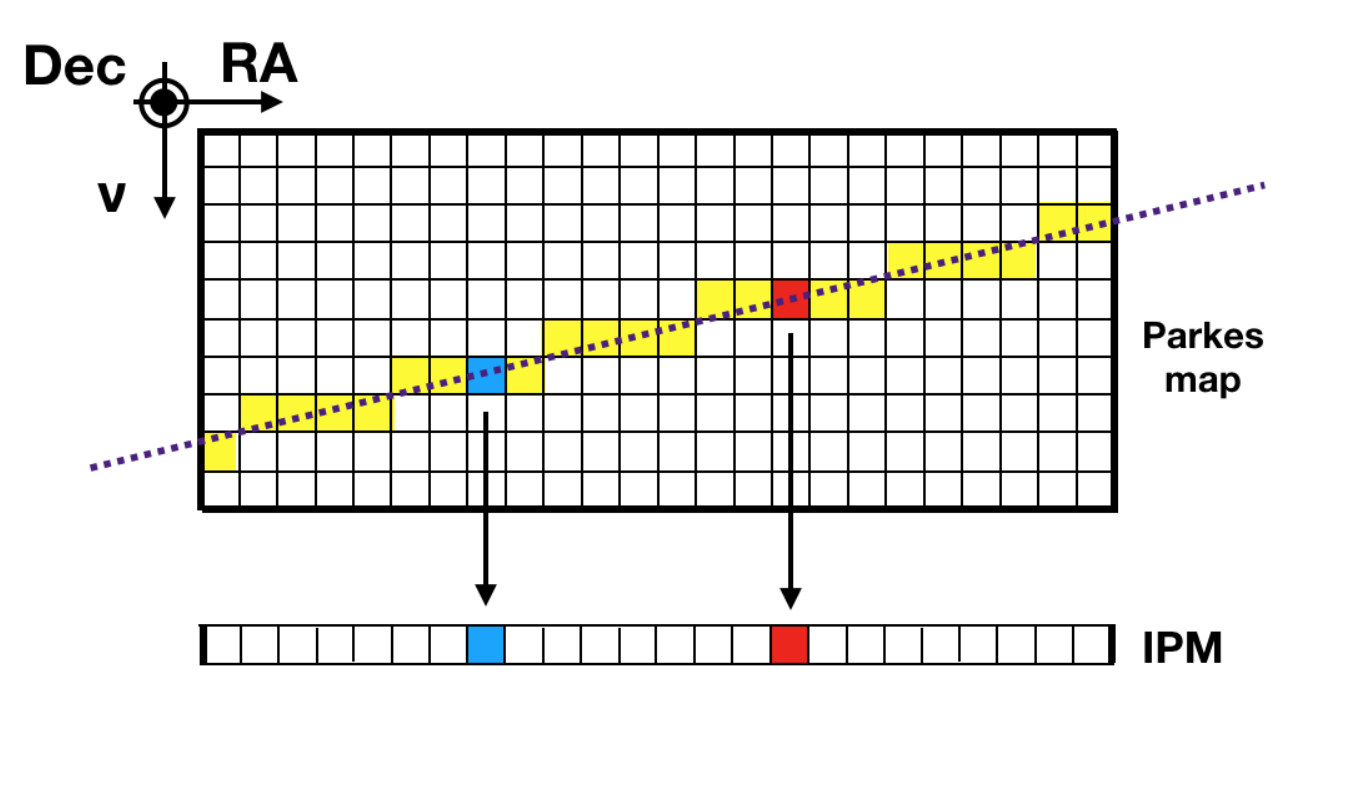}
\includegraphics[trim=  0mm 0mm 10mm 0mm, scale=0.18]{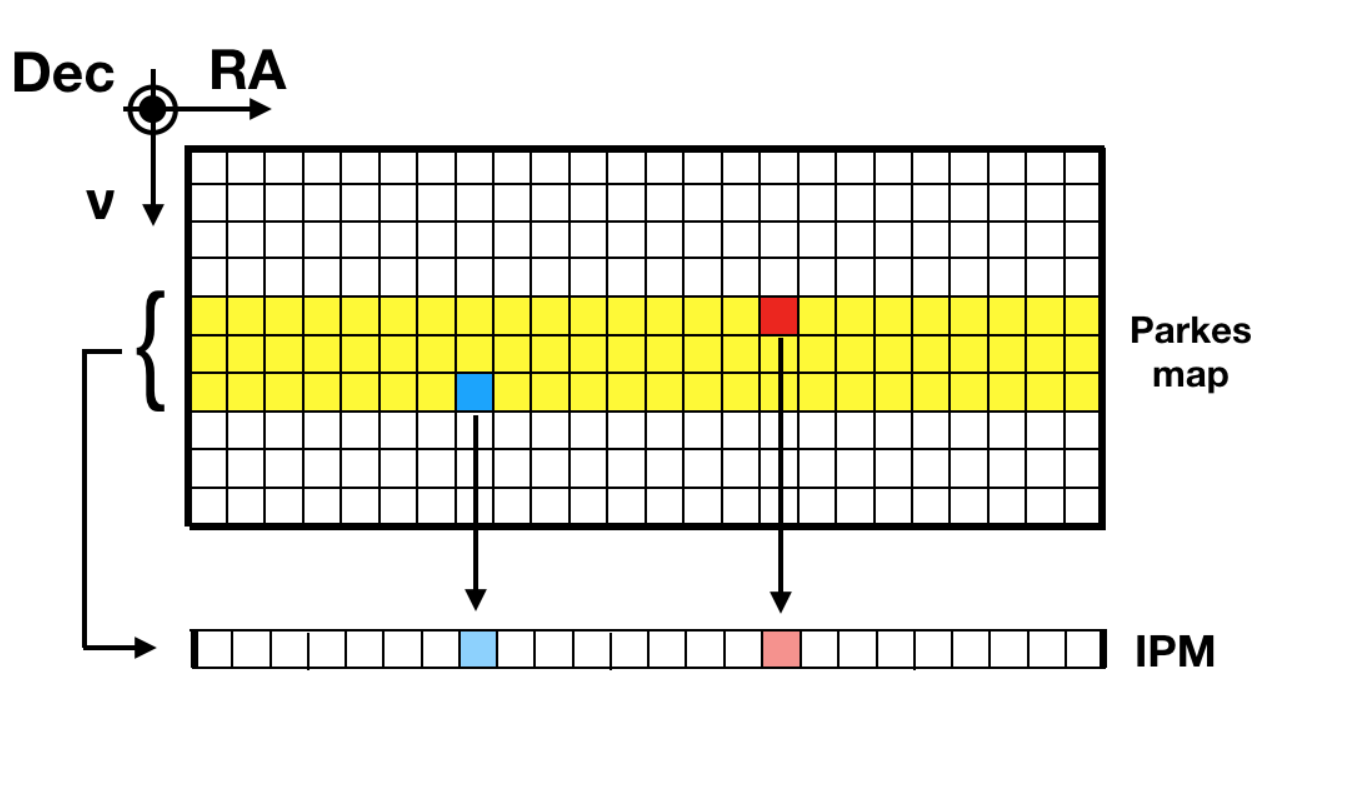}
	\caption{Extraction of an individual pair map 
	(IPM) from the Parkes data cube, shown in this case 
	in projection along the declination axis. The two pixels in red and blue 
	host the galaxies forming the pair, while the pixels highlighted in 
	yellow are the ones defining the IPM.  \textit{Left}. The 
	 method employed in this work is based on the projection along 
	the frequency direction of the pixels intercepted by a plane parallel 
	to the direction of the pair and to either the declination (as in this
	case) or the right-ascension axis, depending on the orientation of the 
	pair in the three-dimensional data structure. This method assumes that 
	the filament is aligned with the pair axis. \textit{Right}. A more conservative 
	average of the frequency/redshift slices spanned by the galaxy pair 
	would have the effect of artificially lowering the intensity of any 
	signal which is not extended across the slices. }
\label{fig:ipm}	
\end{figure*}

The final query we perform on the catalogue aims at selecting the most likely tracers 
 of the nodes in the cosmic web, which typically host massive galaxy clusters with early type  
galaxies. Ideally, one would consider the locally most massive objects, in order to focus 
on the central halo galaxies and discard the effect of satellite galaxies; however, the 2dFGRS catalogue 
does not provide the object masses. We therefore use the galaxy magnitude as a mass proxy, and  
consider the locally brighter galaxies in the R-band. We adopt an isolation criterion similar 
to the one proposed by~\citet{planck13} and also employed by~\citetalias{tanimura19}: 
we discard a given galaxy if a brighter one is found within a tangential distance 
of $1.0\,h^{-1}\text{Mpc}$ and a line-of-sight (LoS) distance of $|c\Delta z| =1000\,\text{km}\,\text{s}^{-1}$.
The online 2dFGRS catalogue provides the apparent R-band magnitude $m_{\text{R}}$ for each object, 
which for a redshift $z$ we convert into absolute magnitude using:
\begin{equation}
	M_{\text{R}} = m_{\text{R}} - 25 - 5\log_{10}{\left[\frac{D_{\text{L}}(z)}{\text{Mpc}}\right]},
\end{equation}
where $D_{\text{L}}(z)$ is the luminosity distance at redshift $z$. 
The effect of this selection on the magnitude and redshift distribution of the 
galaxies is shown in the first two panels of Fig.~\ref{fig:2df}. 
Apart from the overall reduction in the sample size, this cut tends to favour the 
brighter galaxies, which as expected become relatively more abundant, while keeping 
the redshift distribution roughly flat.
The total number of selected galaxies is finally 13,979; the effective number per patch 
is reported in Table~\ref{tab:parkes}.

These catalogues are used to find the pairs of galaxies which are likely to host 
a filament in between. The pair selection adopts a criterion similar to the ones 
used by~\citetalias{tanimura19} and~\citetalias{degraaff19}:
two galaxies are considered to form a valid pair if their tangential separation 
is in between 6 and 14$\,h^{-1}\text{Mpc}$, and their LoS separation is
less than $6\,h^{-1}\text{Mpc}$. 
The resulting pair numbers per patch are reported in Table~\ref{tab:parkes};
overall, the selection yields 274,712 pairs, a number which is comparable to the 
sample size used by~\citetalias{tanimura19}. The resulting redshift distribution for these pairs 
is shown in the third panel of Fig.~\ref{fig:2df}.


\section{Methodology for galaxy-pair stacking}
\label{sec:methodology}

The galaxies in the 2dFGRS pair catalogue described in Section~\ref{ssec:2dfgalaxies} 
represent the endpoints of possible filaments in the LSS; in order for the 
signal coming from the neutral hydrogen component in these filaments to be 
measurable, their emission is to be combined in such a way as to enhance the 
the local filament signal with respect to the background. This is achieved via a proper 
pair stacking which resembles the works of~\citetalias{degraaff19} and~\citetalias{tanimura19}. 
The current analysis, however, 
 presents a major difference: while their stacking was performed 
on the Compton parameter map, which has no depth, in our case the Parkes 
patches are also extended in the frequency/redshift direction. Our 
stacking algorithm requires therefore a prior step to extract a two-dimensional map for each 
individual pair; these maps can then be properly stacked together. Once the final 
stack maps are obtained, the signal coming from the galaxies defining the filament edges
 has to be removed, and from the resulting residual map it is possible to 
 extract the filament signal. The evaluation of the uncertainties on the final filament 
 emission is performed by repeating the stacking on a set of randomised maps and 
 catalogues. We detail each one of these steps in the following. 

\subsection{Individual pair maps}
\label{ssec:ipm}
A general valid galaxy pair is extended along all the three axes of right ascension, 
declination and frequency in the Parkes data cube; the first step consists in obtaining a 
map in two dimensions only, (RA, Dec), which carries the contribution of the pair:
hereafter we will call such a map an ``individual pair map'' (IPM). For each pair two 
IPMs have to be evaluated, one for the HI signal and one for the corresponding weights. 

The method we use is based on the projection of the pixel values along the line of sight 
(see Fig.~\ref{fig:ipm}, left panel, for a graphic representation). Let us consider 
a galaxy pair with ($\text{RA}_1$,$\text{Dec}_1$,$\nu_1$) and ($\text{RA}_2$,$\text{Dec}_2$,$\nu_2$) the coordinates 
of the two member galaxies; we label $\Delta\text{RA}$ and $\Delta\text{Dec}$ the absolute values of the  
separation along the two equatorial coordinates. 
We build the IPM by setting a pixel in the position
(RA,Dec) to the value found in the pixel (RA,Dec,$\nu_{\rm proj}$) in the Parkes data cube, 
with the projection frequency $\nu_{\text{proj}}$ defined as:
\begin{equation}
	\nu_{\text{proj}}(\xi) = \nu_1 + \dfrac{\nu_2 - \nu_1}{\xi_2-\xi_1} (\xi-\xi_1).
\end{equation}
where the generic variable $\xi$ stands for either RA or Dec. If $\Delta\text{RA}\geq\Delta\text{Dec}$, then 
$\xi=\text{RA}$ and the value of $\nu_{\text{proj}}$ is the same for all the pixels in the IPM with the same right 
ascension; conversely, if $\Delta\text{RA}<\Delta\text{Dec}$, then 
$\xi=\text{Dec}$ and the value of $\nu_{\text{proj}}$ is the same for all the pixels in the IPM with the same declination.
In other words, the IPM is built by projecting along the frequency direction a tilted plane 
defined by the separation vector and either the declination or the right ascension versor, depending on 
whether the galaxies are separated mostly in RA or Dec. 
 By construction, this method preserves the values of the HI signal stored in the initial map;
in particular, it provides an IPM which reports the correct amplitude for all the pixels 
on the line that joins the two galaxies in the Parkes data cube. 
For pairs located near the edges of the 
frequency band it is possible that for values of RA or Dec far from the pair centre the 
resulting value of $\nu_{\text{proj}}$ falls out of the allowed frequency range; 
in this case the affected IPM pixels are properly flagged and their values in the 
weight IPM is set to zero. 
\begin{figure*}
\includegraphics[trim= 0mm 0mm 0mm 0mm, scale=0.36]{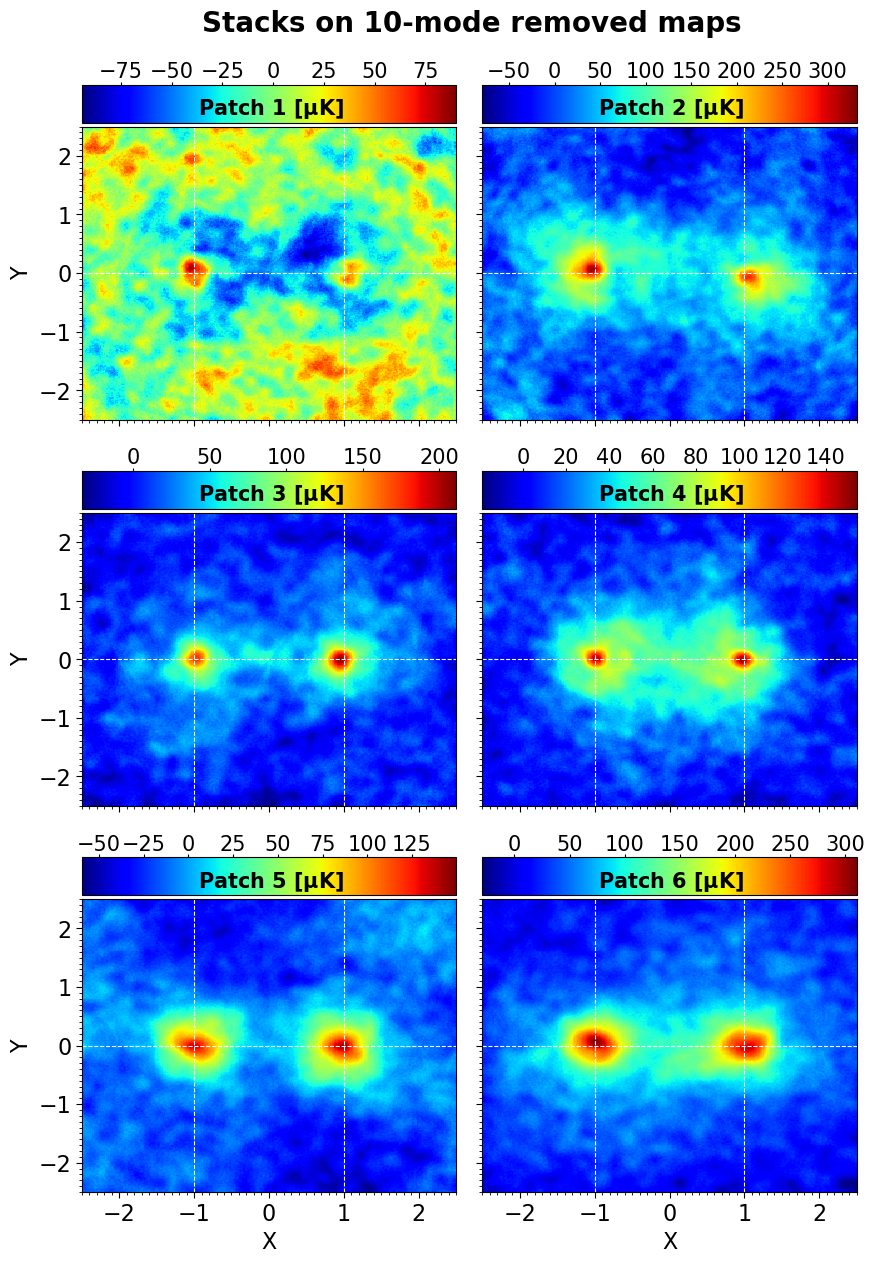}\qquad
\includegraphics[trim= 0mm 0mm 0mm 0mm, scale=0.36]{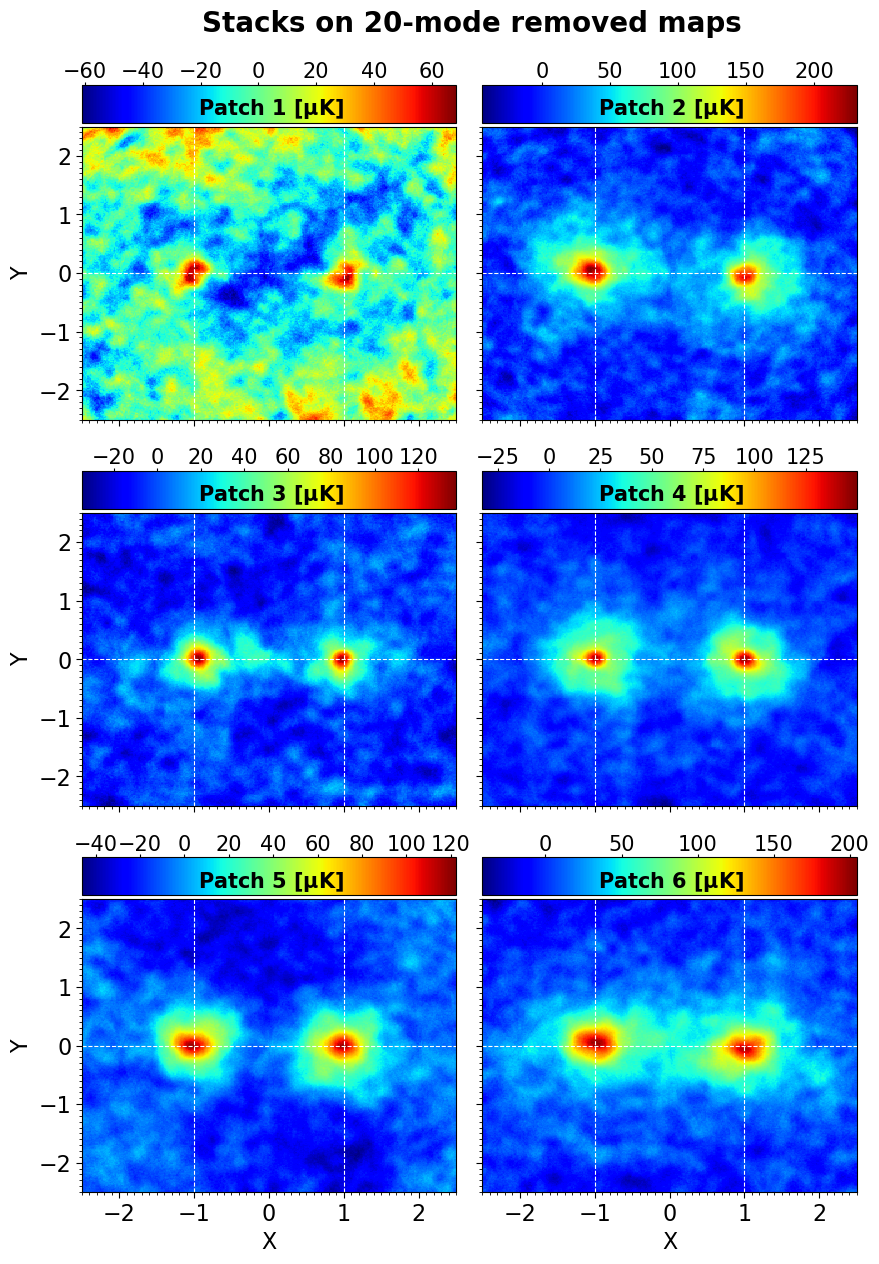}
\caption{Final maps obtained by stacking the selected 2dFGRS galaxy pairs 
	on the Parkes 21-cm maps, for the case of 10 (\textit{left}) and 20
	(\textit{right}) modes removal. We show the results for all six patches; in order to enhance 
	the details of these maps we chose to saturate the color scale for each patch, 
	resulting in different ranges. With the exception of patch 1, which is apparently noisier, 
	the rest of the stacks show similar patterns. The two peaks corresponding to the stacked 
	2dFGRS galaxies are clearly visible in their expected positions, but at this level the 
	possible presence of a linking filament is not clear. }
\label{fig:stack_all}	
\end{figure*}

The method we have described implicitly assumes that the filaments connecting the 
galaxy pairs are straight lines. Any strong curvature of the filament 
would make it deviate from 
the segment joining the two endpoints and bring it out of the projection plane, meaning that 
it would not enter the corresponding IPM. We also explored a different method to work around 
this issue, based on the average of the contribution from all the frequency slices located 
in between the two galaxies (Fig.~\ref{fig:ipm}, right panel).
This method allows to mitigate the loss of bent filaments, provided their curvature is not so large that 
it takes them outside the frequency range bounded by the two galaxies. 
However, this approach produces an artificial dilution of any signal that is not coherently repeated across slices along 
the line of sight (including the 
filament signal we are looking for), the effect being higher for larger redshift separations of the two galaxies.  
Given the complex spatial behaviour of both the signal and the weights in the three-dimensional 
Parkes patches, it is not possible to quantify the effect with a simple scaling factor to be applied in 
the end as a correction. Furthermore, we should notice that features that extend along 
the frequency direction and are enhanced by this method are usually foreground 
residuals, so the final IPM is likely to be 
affected by a higher foreground contamination.

For this work we decided to use the projection method for building the IPMs. 
Indeed, even though some filaments may lie outside the geometrical line 
joining their endpoints, the majority of the pair galaxies are separated 
by only one or two slices: while this is enough to dilute any filament signal 
by a factor 2 or 3 using the averaging method, it is unlikely that the filament 
spine avoids the totality of the pixels intercepted by the projection plane. 
In addition, the frequency resolution 
of the Parkes maps translates into a slice thickness of $\sim 2\,\text{Mpc}$, 
which is the typical expected diameter for a filament: although in some sections 
of the projection plane the filament spine may lie in the frequency slice right 
next to the one that is being considered, the filament outskirts are expected to 
enter the IPM in the projection. 
In any case, this is an approximation that undoubtedly affects 
the results of a filament blind search like the one we are 
undertaking in this work, and has to be acknowledged.

\begin{figure*}
\includegraphics[trim= 0mm 0mm 0mm 0mm, scale=0.19]{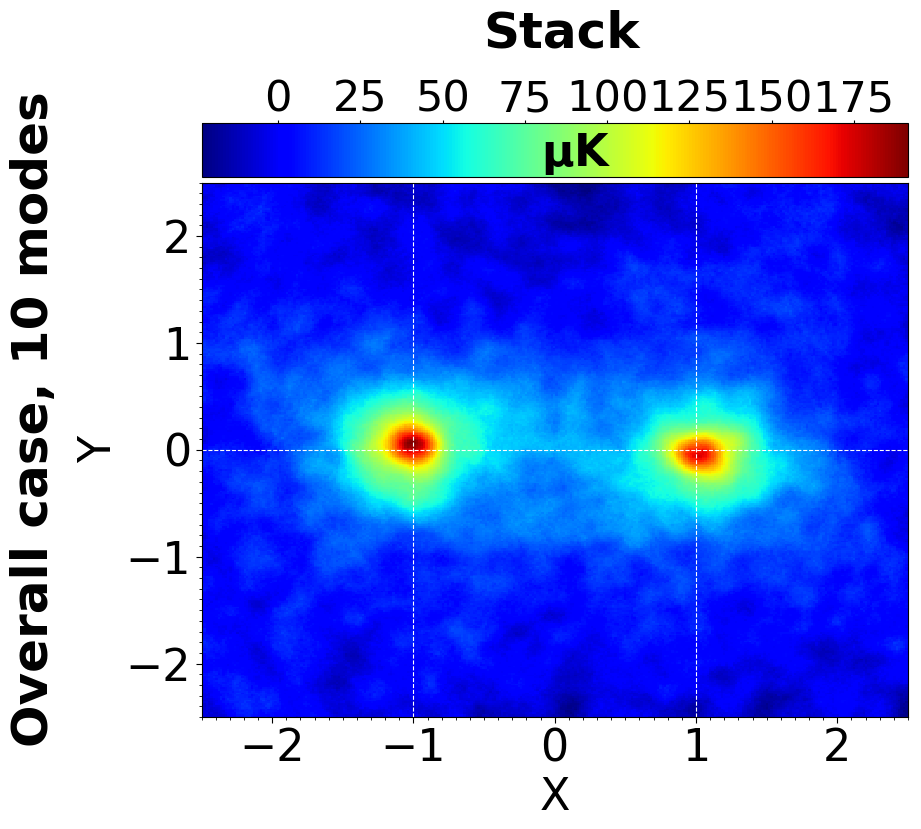}
\includegraphics[trim= 0mm 0mm 0mm 0mm, scale=0.19]{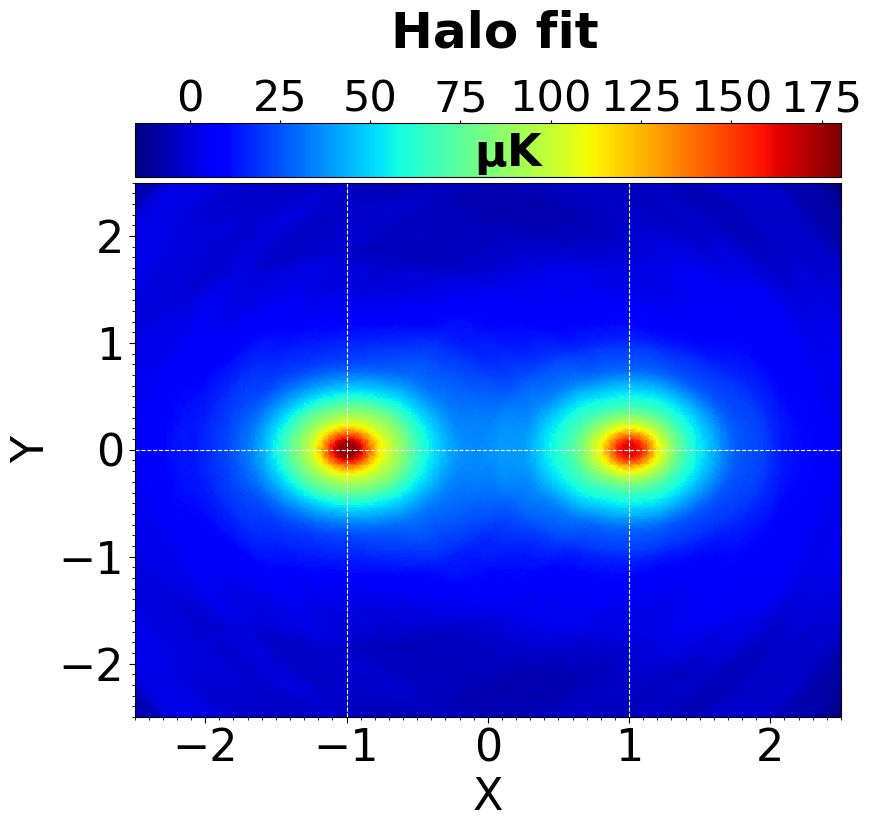}
\includegraphics[trim= 0mm 0mm 0mm 0mm, scale=0.19]{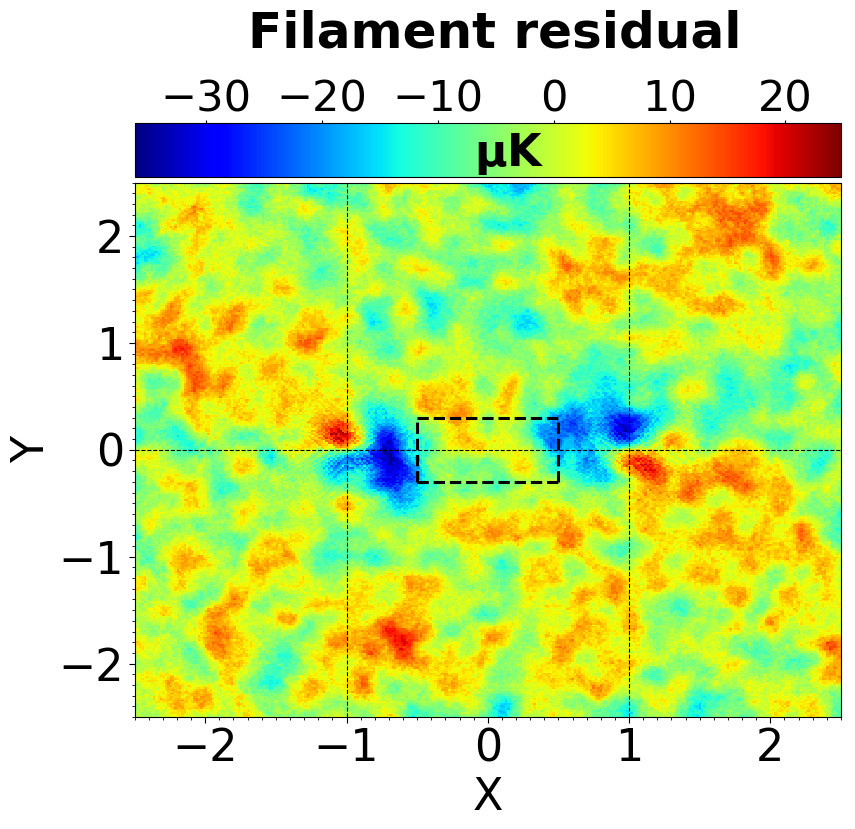}
\includegraphics[trim= 0mm 0mm 0mm 0mm, scale=0.16]{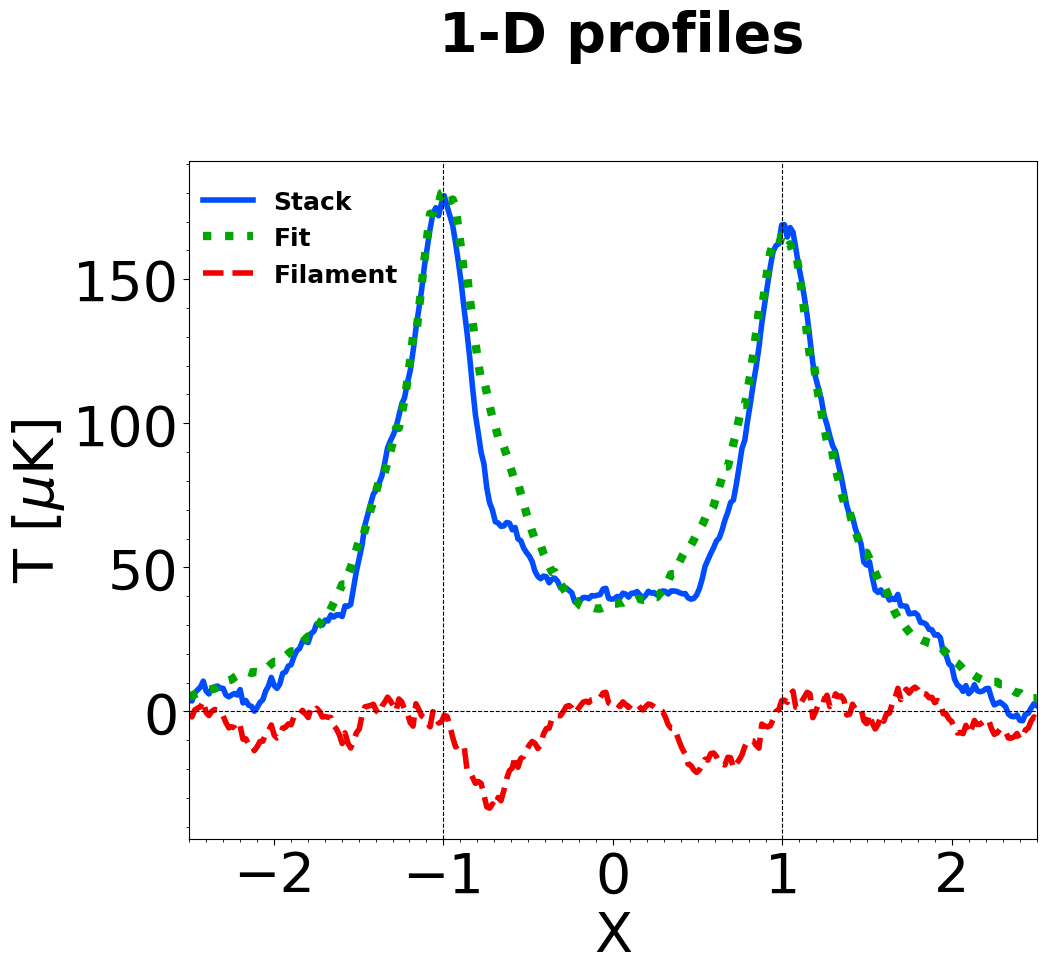}\newline
\includegraphics[trim= 0mm 0mm 0mm 0mm, scale=0.19]{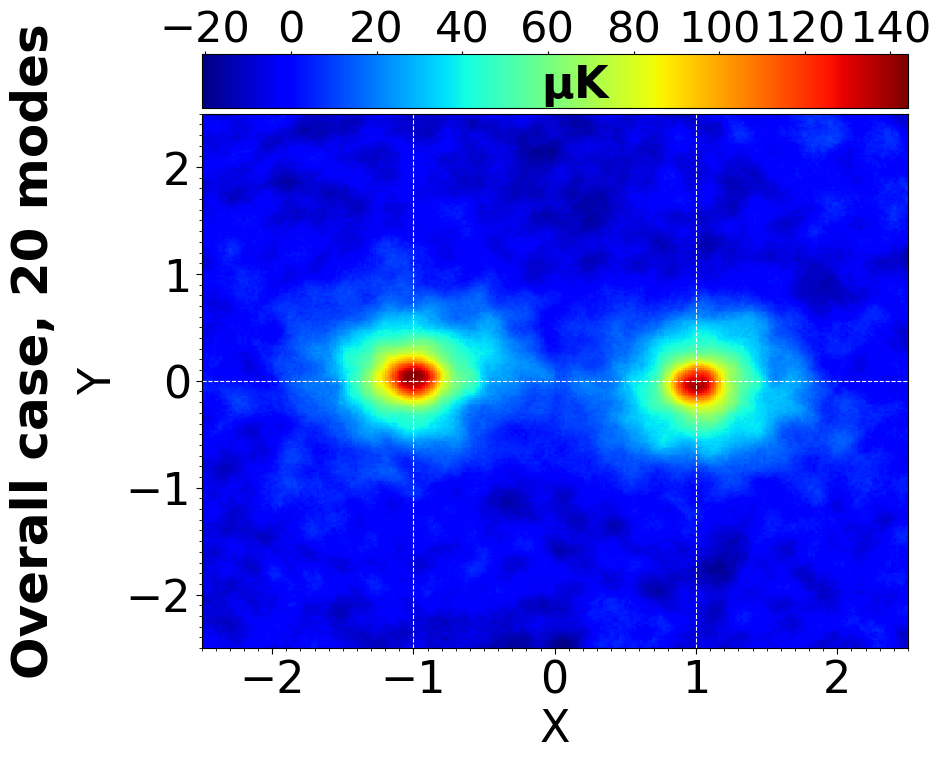}
\includegraphics[trim= 0mm 0mm 0mm 0mm, scale=0.19]{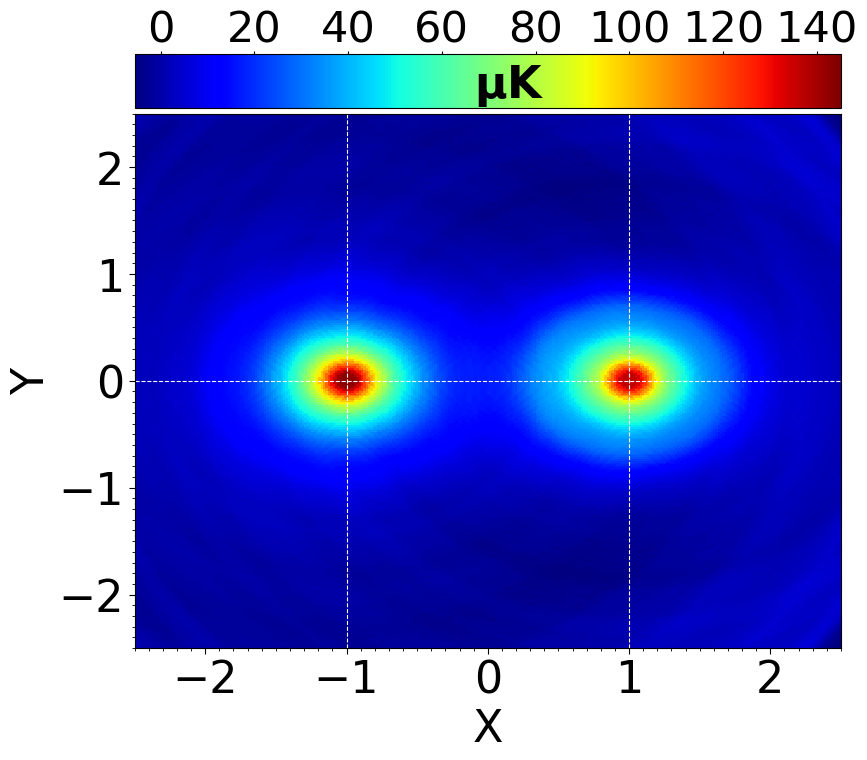}
\includegraphics[trim= 0mm 0mm 0mm 0mm, scale=0.19]{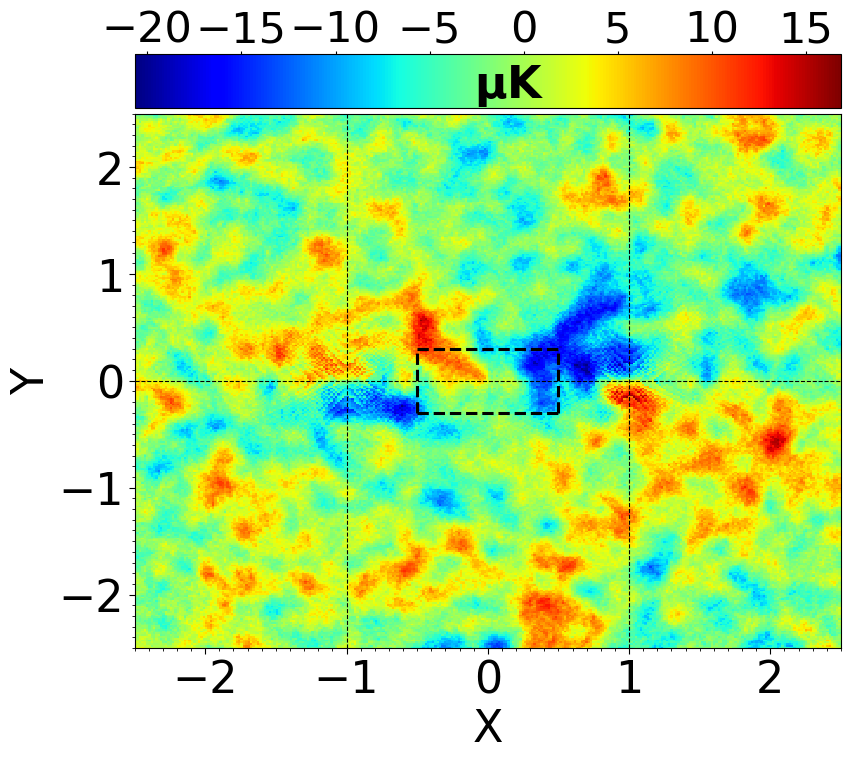}
\includegraphics[trim= 0mm 0mm 0mm 0mm, scale=0.16]{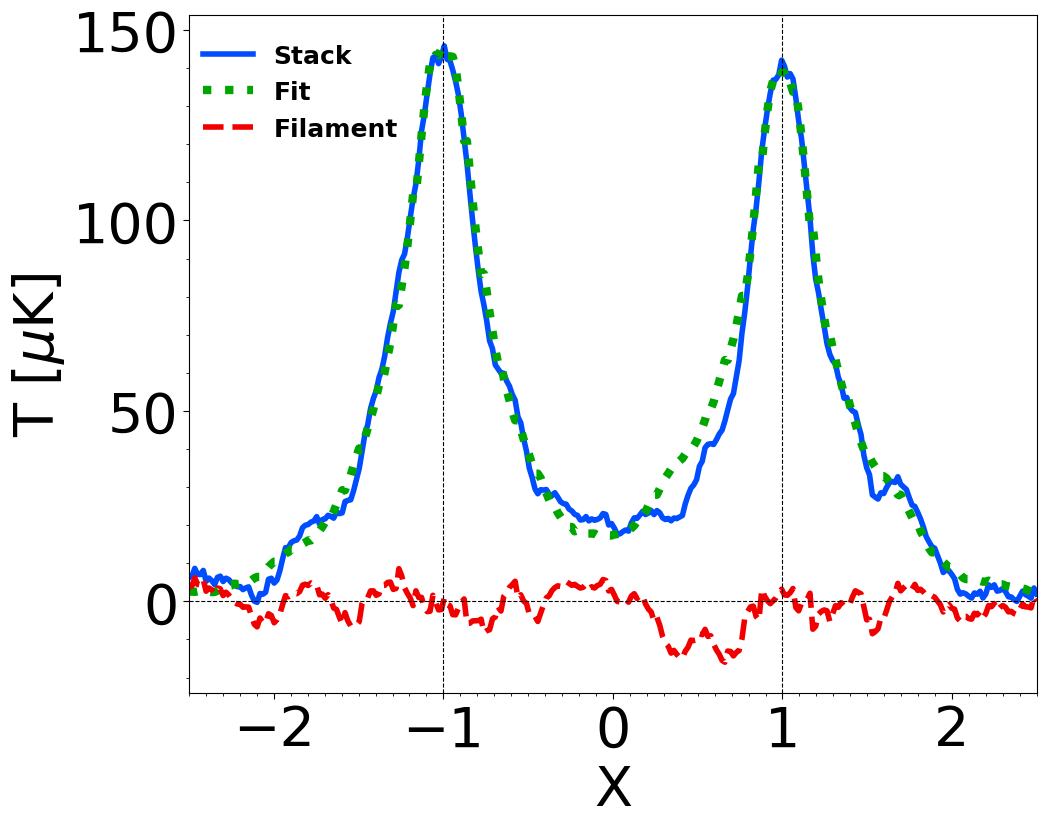}
\caption{Results for the analysis of the full map obtained by combining the contribution of 
	all patches (see eq.~\eqref{eq:map_tot} in the text), shown for both the 10-mode (\textit{Top row})
	and the 20-mode (\textit{bottom row}) removed maps. 
	\textit{First column.} Overall stacking result map; the 
	improvement of the statistics deriving from combining the individual patch contributions 
	determines a more homogeneous background in this final stack. 
	\textit{Second column.} The corresponding halo fit map, 
	carrying the contribution from the galactic HI peaks only. 
	\textit{Third column.} The filament residual 
	map, obtained by subtracting the halo-fit map from the stack map. 
	\textit{Fourth column.} The one dimensional 
	profiles at Y=0 for the total stack, the halo fit and the filament residuals, confirming the 
	absence of a net filament excess in the central region. }
\label{fig:stack_total}	
\end{figure*}

\subsection{Stacking procedure}
\label{ssec:stacking}

The galaxy pairs we consider have different positions, angular separations and 
orientations in the IPM maps. 
In order to make sure that the signal from the pair galaxies and the filament in between 
is added consistently, we have to bring all pairs to a common orientation. 
To this aim, we define a new reference frame (X,Y), with origin in the 
pair central coordinates and whose axes are defined in such a way that one of the
galaxies lies in the position $(-1,0)$ and the other in the position $(+1,0)$. 
Each IPM is therefore brought to this common reference frame, by applying a 
rotation around the pair centre that brings the galaxies along the 
X-axis, and by rescaling the lengths by a common scale factor in order to place the 
galaxies at unit distance from the centre.
This procedure is applied to both the signal and the weight IPMs, generating 
a set of final maps $S_{\text{XY}}^{i}$ and $W_{\text{XY}}^{i}$ for 
 $i=1,...,N_{\rm p}$, with $N_{\rm p}$ the total number of pairs. The final stacked map for a given patch $n$ is then obtained as:
\begin{equation}
	S_{n} = W_{n}^{-1}\,\sum_{i=1}^{N_{\text{p},n}}S_{\text{XY}}^{i}W_{\text{XY}}^{i},
\end{equation}
where $N_{\text{p},n}$ is the total number of pairs for the patch and
\begin{equation}
	W_{n} = \sum_{i=1}^{N_{\text{p},n}}W_{\text{XY}}^{i}
\end{equation}
is the corresponding total weight. The resulting stack maps for the individual 
patches are shown in Fig.~\ref{fig:stack_all}. We also consider the total 
stack map $S_{\rm T}$ obtained combining the results from individual patches. 
This is done again via a weighted average:
\begin{equation}
	\label{eq:map_tot}
	S_{\text{T}} = \dfrac{ \sum_{n=1}^{6} S_{n} W_{n} }{\sum_{n=1}^{6} W_{n}}.
\end{equation}
The result is shown in the leftmost panels of Fig.~\ref{fig:stack_total}; 
hereafter we shall refer to this case as the overall stack.

\subsection{Filament extraction}
\label{ssec:filament}

The final stack maps clearly show two peaks located in the expected 
positions $(\text{X},\text{Y})=(\pm1,0)$. This emission proceeds 
from the HI bulk located inside the galaxies that 
we are stacking; since in our data set these galaxies have been selected to locate
the LSS halos, we can refer to each of these peaks as the one-halo emission. 
This emission is clearly the dominant feature in our stacks, and any 
filament signal will be a second order contribution. In order to assess 
quantitatively the presence of actual filaments in the maps
we have to remove the emission from the pair galaxies. 

First of all, since we are interested in the relative emission excess with 
respect to the background, we evaluate the mean background level as the average 
of the pixels found in the external frame defined by the conditions 
$|\text{X}|>2$ or $|\text{Y}|>1.5$, and subtract it from each map as 
an overall offset. Secondly, for each stack we build a map carrying 
only the contribution from the 
halo peaks, obtained by fitting a halo profile on the
 stack maps. Notice that in this case it is not meaningful to 
 derive an analytical expression to fit for; indeed, the one-halo profiles visible 
in the stacks are obtained from the superposition of several profiles 
from individual galaxies with different masses, each of which underwent a 
different rescaling during the processing of the corresponding IPM (depending 
on the initial pair separation in the (RA,Dec) frame, 
the profiles are either shrinked or 
stretched in order to match the final separation of $\Delta$X=$2$ in 
the (X,Y) frame). This implies that 
an analytical expression for the radial dependence of the HI intensity 
in these maps will be uncorrelated from the physical parameters describing the 
stacked halos (like their mass, brightness temperature or redshift). 
\begin{figure*}
\includegraphics[trim= 0mm 0mm 0mm 0mm, scale=0.37]{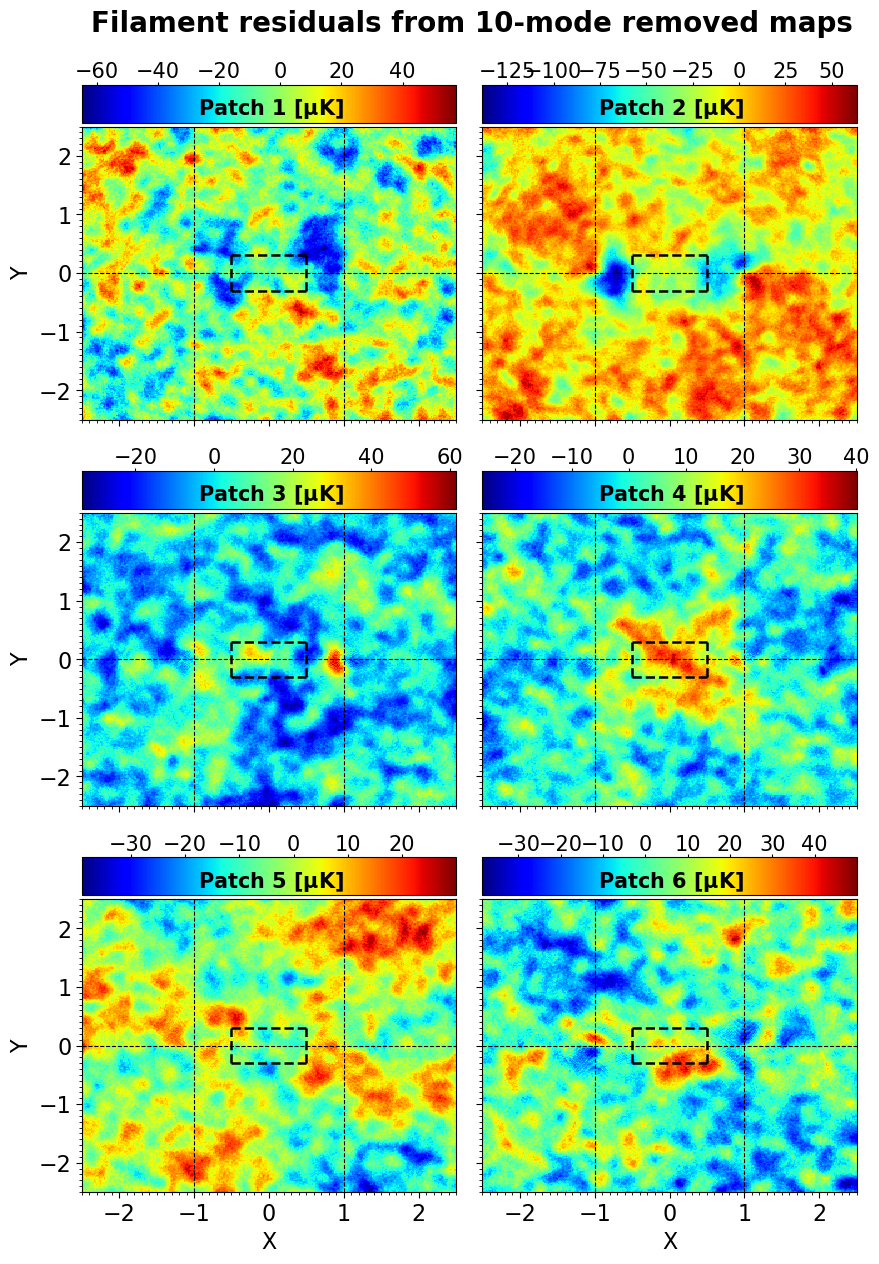}\qquad
\includegraphics[trim= 0mm 0mm 0mm 0mm, scale=0.37]{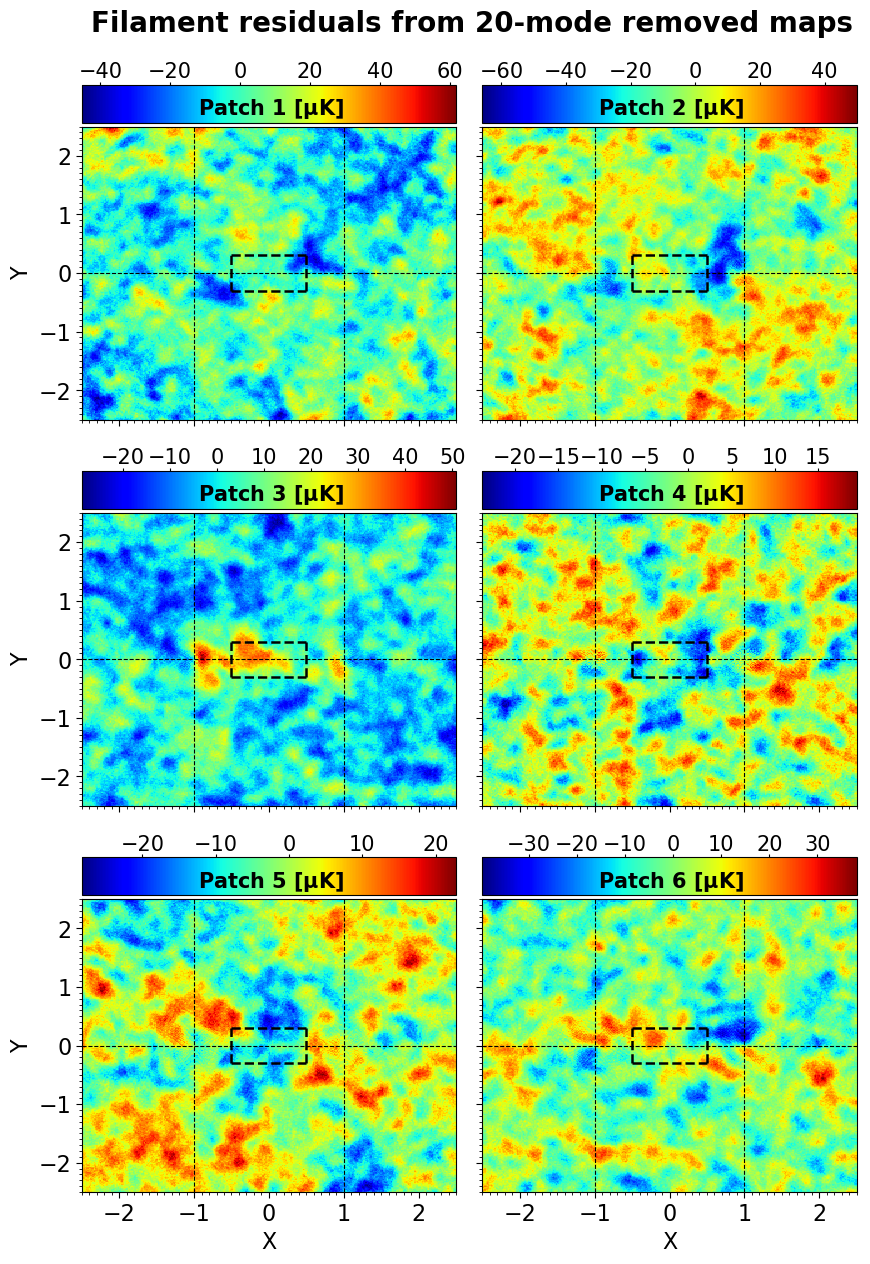}
\caption{Filament residual maps for the six patches, 
	obtained by subtracting from the stack maps in Fig.~\ref{fig:stack_all} the corresponding 
	fitted halo profiles. Again, the scale is saturated for each individual patch in order 
	to better show the background structures.}
\label{fig:filament_all}	
\end{figure*}

We therefore use a different approach and extract the profiles 
directly from the map, with the only assumption that 
the profile for each halo has to be circularly symmetric. 
For the peak centred 
in X=-1 we define a set of $N_{\rm b}$ bins $r_i$ 
(with $i=1, ..., N_{\rm b} $) for the radial distance from the 
point $(-1,0)$ with separation $\Delta r = 0.02$ in the scaled units; 
for a generic profile $P_i$ defined over these bins we can build a 
one-peak map $M_{-1}$ by assigning the value $P_i$ to all pixels with distance 
from the position $(-1,0)$ in the range $[r_i, r_i+\Delta r]$. We can 
repeat the same procedure for the other peak centred in $(+1,0)$ and 
for a chosen profile $P_j$ we can build the corresponding map $M_{+1}$.
The sum of the two one-peak maps, $M_{\text{peaks}}=M_{-1}+M_{+1}$, is our 
ansatz representing the halo-only contribution to the stack map. 
The final, best fit estimate of such a map is obtained by varying the
one-halo profiles $P_i$ and $P_j$, and looking for the set of values that 
minimises the squared difference between the stack map and the peaks map.
Since the maps are not perfectly symmetric, we allow the two halo peaks 
to have different radial profiles; plus, by fitting the sum of the two one-halo 
peak maps we also take into account the contribution of each galaxy profile on the
other. In order for the final profiles to be representative of the halo 
contribution only, we restrict our fit to the external region of the map, 
in a frame defined by the conditions $|\text{X}|>1$ or $|\text{Y}|>1$. 
This way we ensure that the profile fit is not affected by any spurious filament 
contribution. An example of the resulting fit map, for the case of the overall 
stack, is shown in the second column panels of 
Fig.~\ref{fig:stack_total} (the fit maps for the 
individual patch cases are qualitatively very similar to these ones and do 
not provide any further information).
\begin{figure*}
\includegraphics[trim= 0mm 0mm 0mm 0mm, scale=0.32]{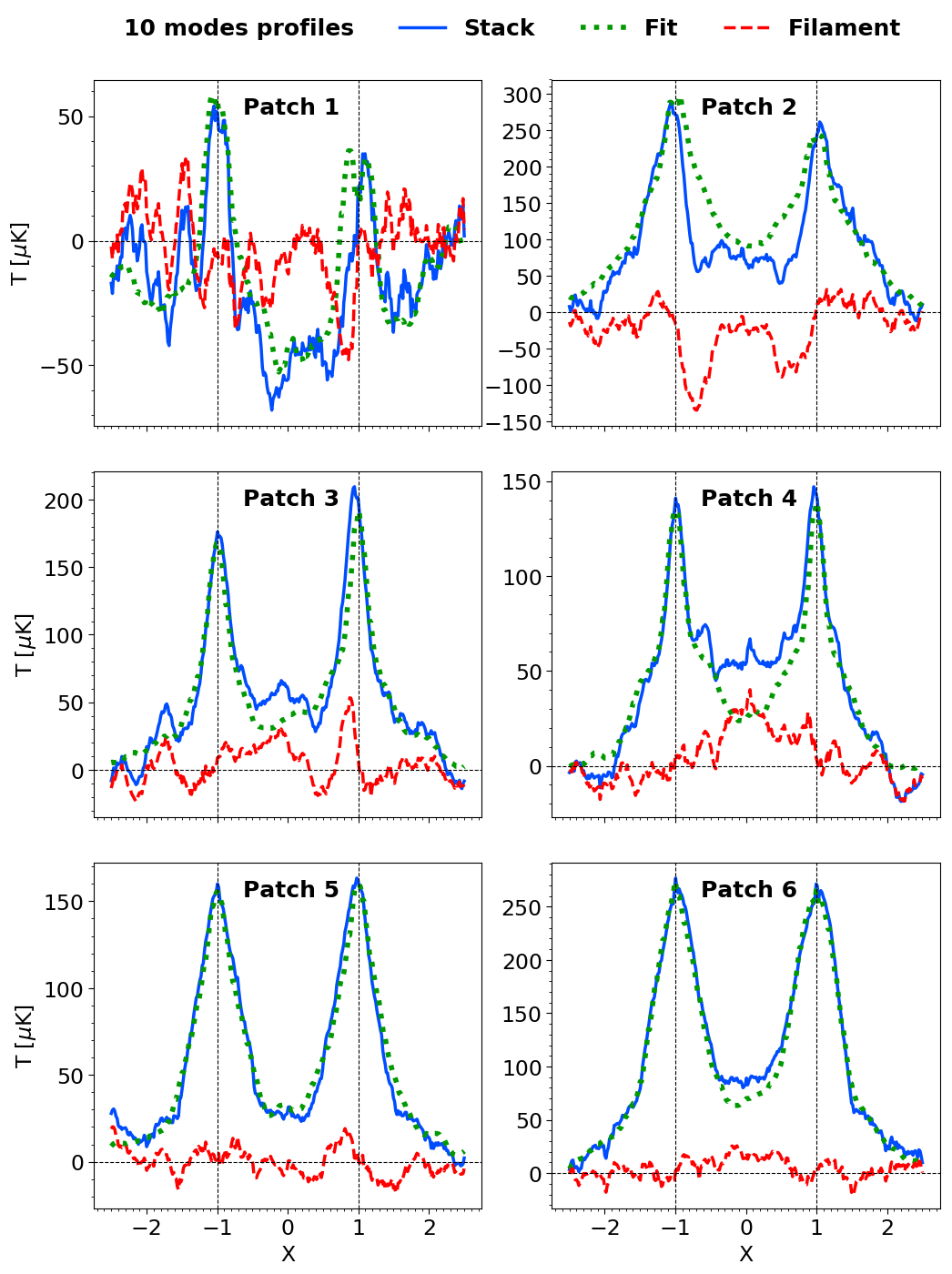}\qquad
\includegraphics[trim= 0mm 0mm 0mm 0mm, scale=0.32]{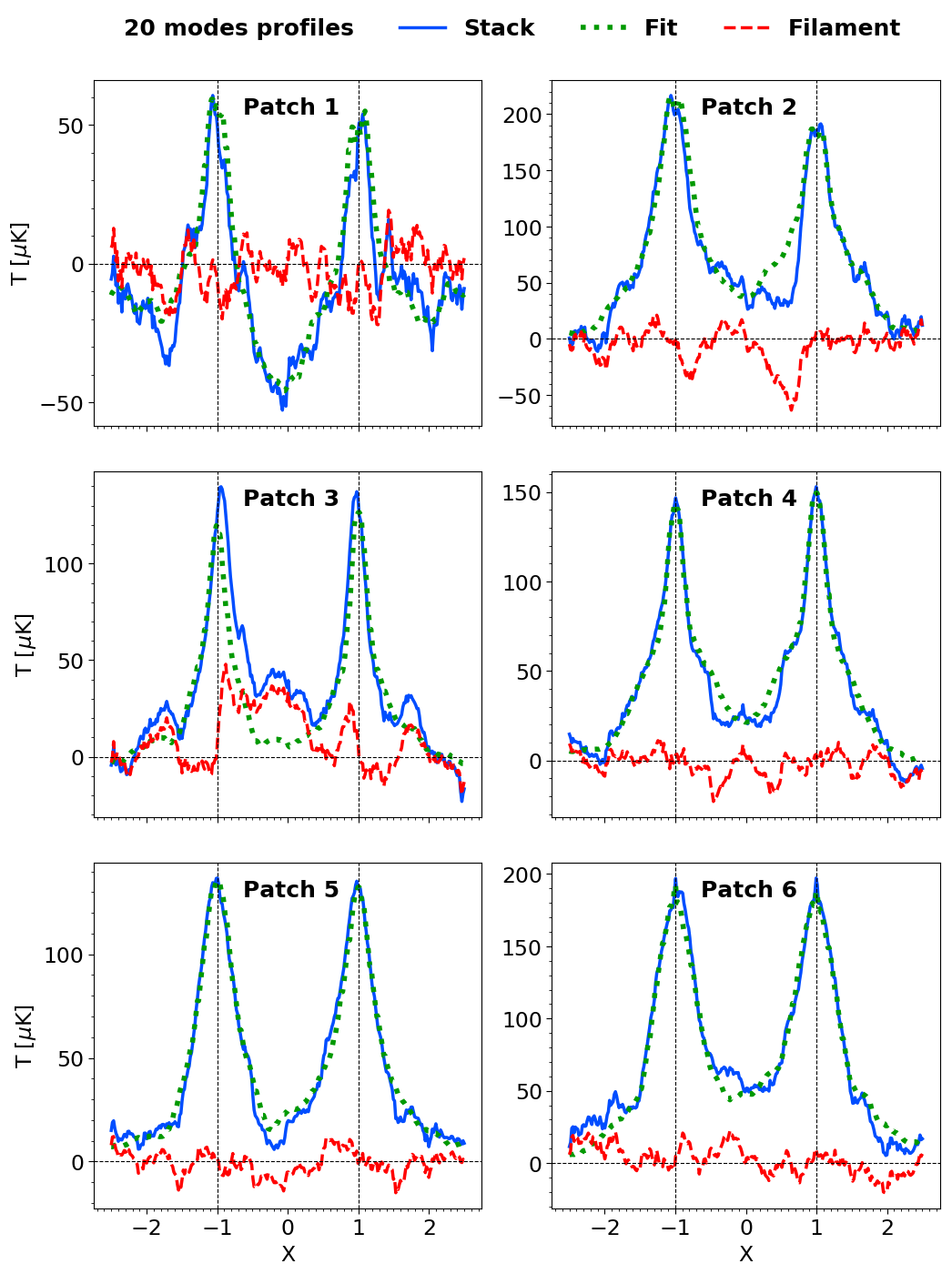}
\caption{ One dimensional cuts 
	at Y=0 of the stacked maps from Fig.~\ref{fig:stack_all}, the residual maps from Fig.~\ref{fig:filament_all}
	and the associated fit maps (not shown), for both the 10-mode (\textit{left}) and the 
	20-mode (\textit{right}) removal cases.
	Generally, the sum of the two individual profiles is enough to account for 
	the signal visible in the central area, leaving no room for a filament contribution.}
\label{fig:profiles_all}	
\end{figure*}

At this point, the subtraction between the stack map and the fitted halo 
map will be free from the galaxy contribution and show any possible filament 
emission in terms of an excess of signal in the central region. 
Fig.~\ref{fig:filament_all} shows the resulting filament residual map for each 
patch, while the third column panels in Fig.~\ref{fig:stack_total} show 
the filament maps for the overall stacks. The relative contributions from 
the stack and the fit are better visualised in terms of one-dimensional 
cuts of these maps; the corresponding profiles as a function of 
X at Y=0 (that is, along the galaxy pair axis), together with the residual 
filament contribution, are shown 
in Fig.~\ref{fig:profiles_all} for the individual patches and in the 
fourth column panels of Fig.~\ref{fig:stack_total} for the overall stacks. 
However, the measurement of the filament emission is not performed on these 
cuts, but on the two-dimensional residual maps. 
The filament emission intensity is estimated as the mean of the map pixels found in 
the region $|\text{X}|<0.5$, $|\text{Y}|<0.3$, which is shown as a dashed box in the residual maps;
the resulting values are plotted in Fig.~\ref{fig:summary_fil} and will be 
discussed in Section~\ref{sec:discussion_stack}.

\subsection{Error estimation}
\label{ssec:randstack}

In order to establish the significance for the filament detections we 
estimate their uncertainties adopting a bootstrap approach, based on the 
repetition of the stacking using a randomised data set. We 
consider two different implementations of this method. 
\begin{figure*}
\includegraphics[trim= 0mm 0mm 0mm 0mm, scale=0.37]{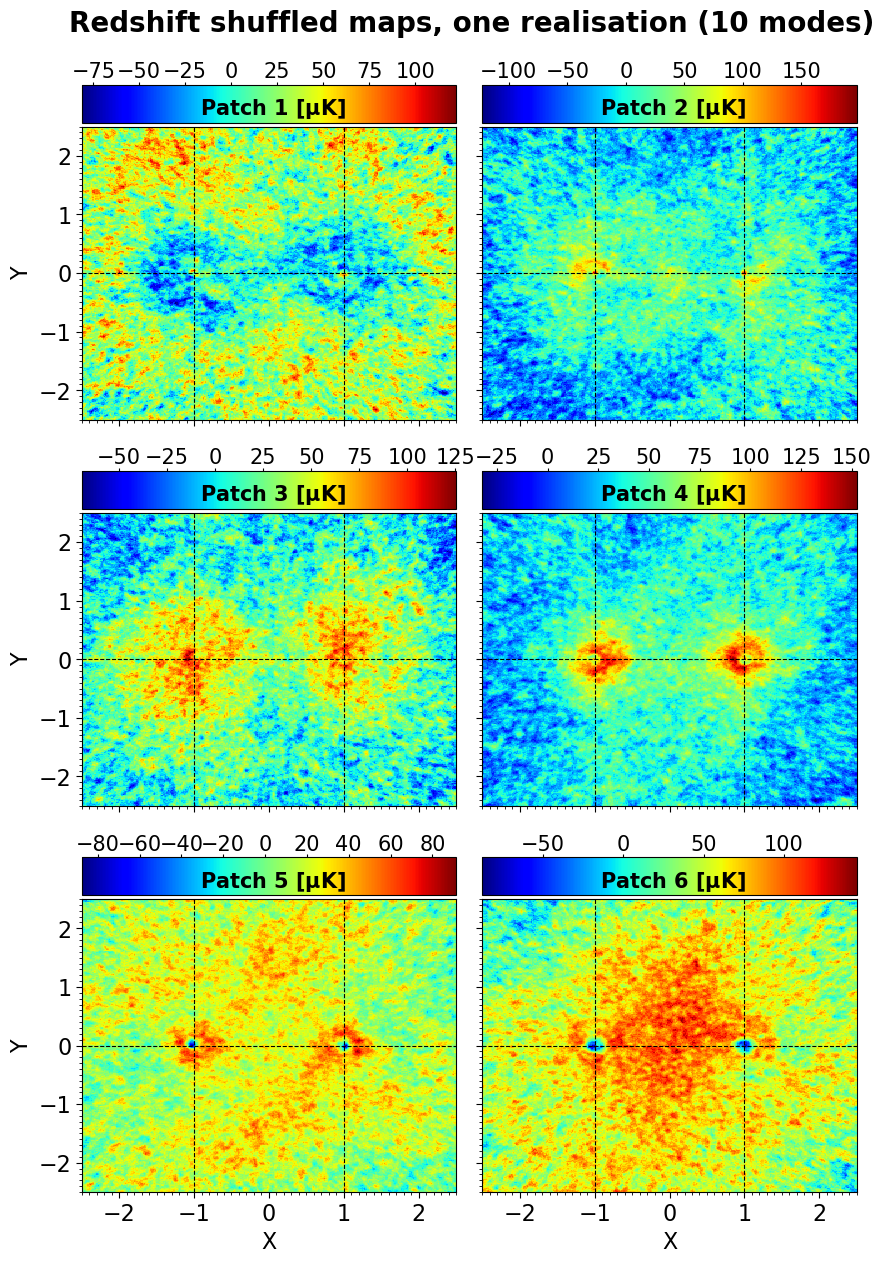}\qquad
\includegraphics[trim= 0mm 0mm 0mm 0mm, scale=0.37]{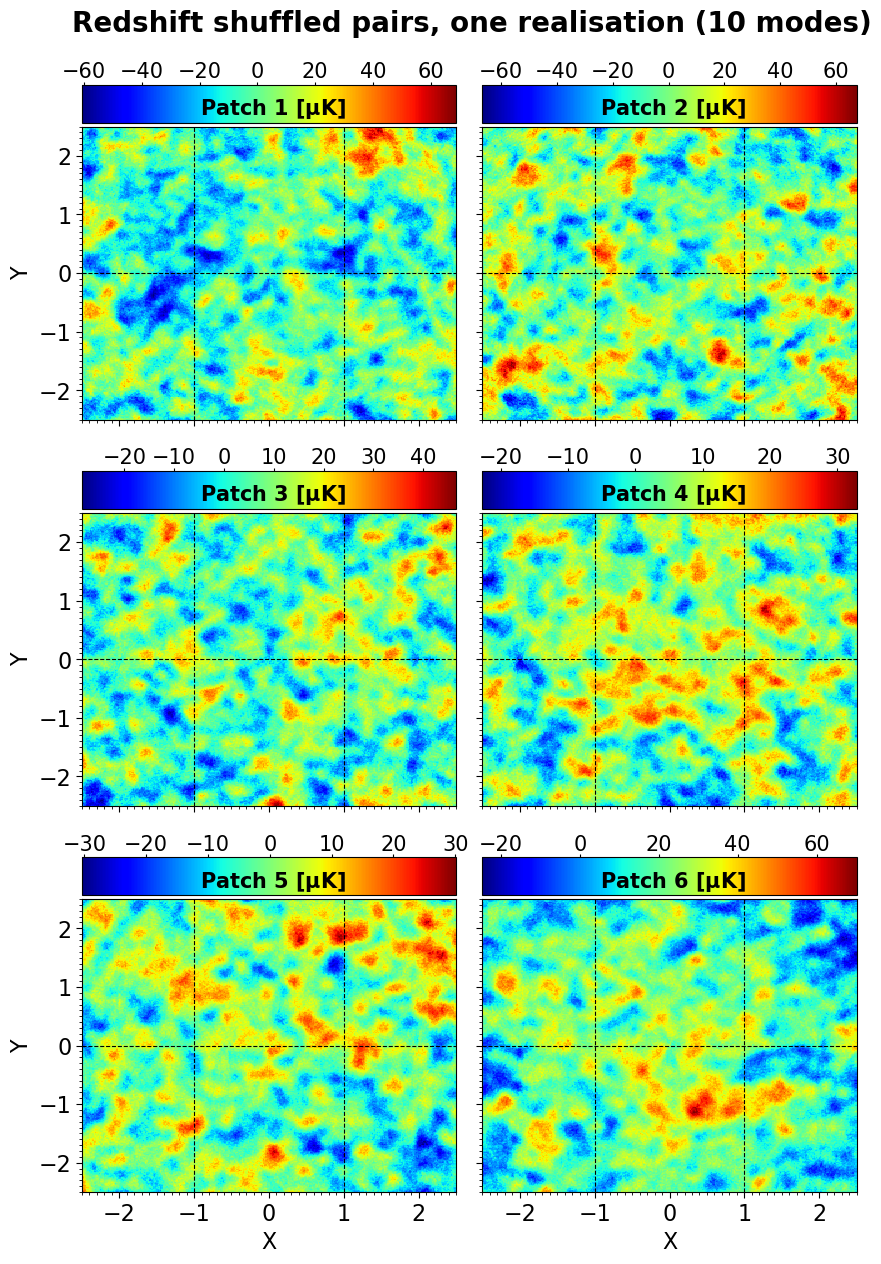}
	\caption{Results per patch of the two implementation of the bootstrap randomisation, shown for one 
	particular realisation out of the total 500 for the 10-mode removed maps. \textit{Left.}
	Stacking of the original pair catalogue on the Parkes maps with shuffled redshift slices. \textit{Right.} 
	Stack of the pair catalogue with shuffled redshifts on the original Parkes maps. }
\label{fig:rand_realis}	
\end{figure*}

In the first case, we keep the pair catalogue unchanged and randomise the Parkes maps. 
The randomisation of individual pixel positions in the (RA, Dec) frame would not be appropriate because it would 
disrupt the correlation between adjacent pixels introduced by the Parkes beam, and 
create artificial structures with typical extension below the size of the main beam. 
We therefore shuffle the map redshifts: each frequency/redshift slice map is kept unchanged, but 
assigned to a different slice position, generated randomly; it is imposed that no slice can 
end up in its initial position after this shuffling, ensuring that all slices actually change their
place. The same, new slice ordering is applied to all the six Parkes patches. 
We generate this way a set of 500 randomised Parkes maps, and for each one of 
them we repeat the same stacking procedure and filament extraction described 
for the case of the real data. 

The second way we implement the bootstrap method is by keeping the maps 
unchanged and by randomising the pair catalogue. Again, we keep the 
(RA, Dec) coordinates of all pair galaxies fixed and only randomise their 
redshifts. Only the average redshift of each pair is changed, 
but not the redshift separation of the two member galaxies; the new average 
pair redshift is generated as a random number uniformly distributed in 
the range allowed by the Parkes maps. 
We generate this way 500 shuffled pair catalogues, each of which is 
stacked over the real Parkes data to obtain an estimation of the 
filament for each patch.
\begin{figure*}
\includegraphics[trim= 0mm 0mm 0mm 0mm, scale=0.31]{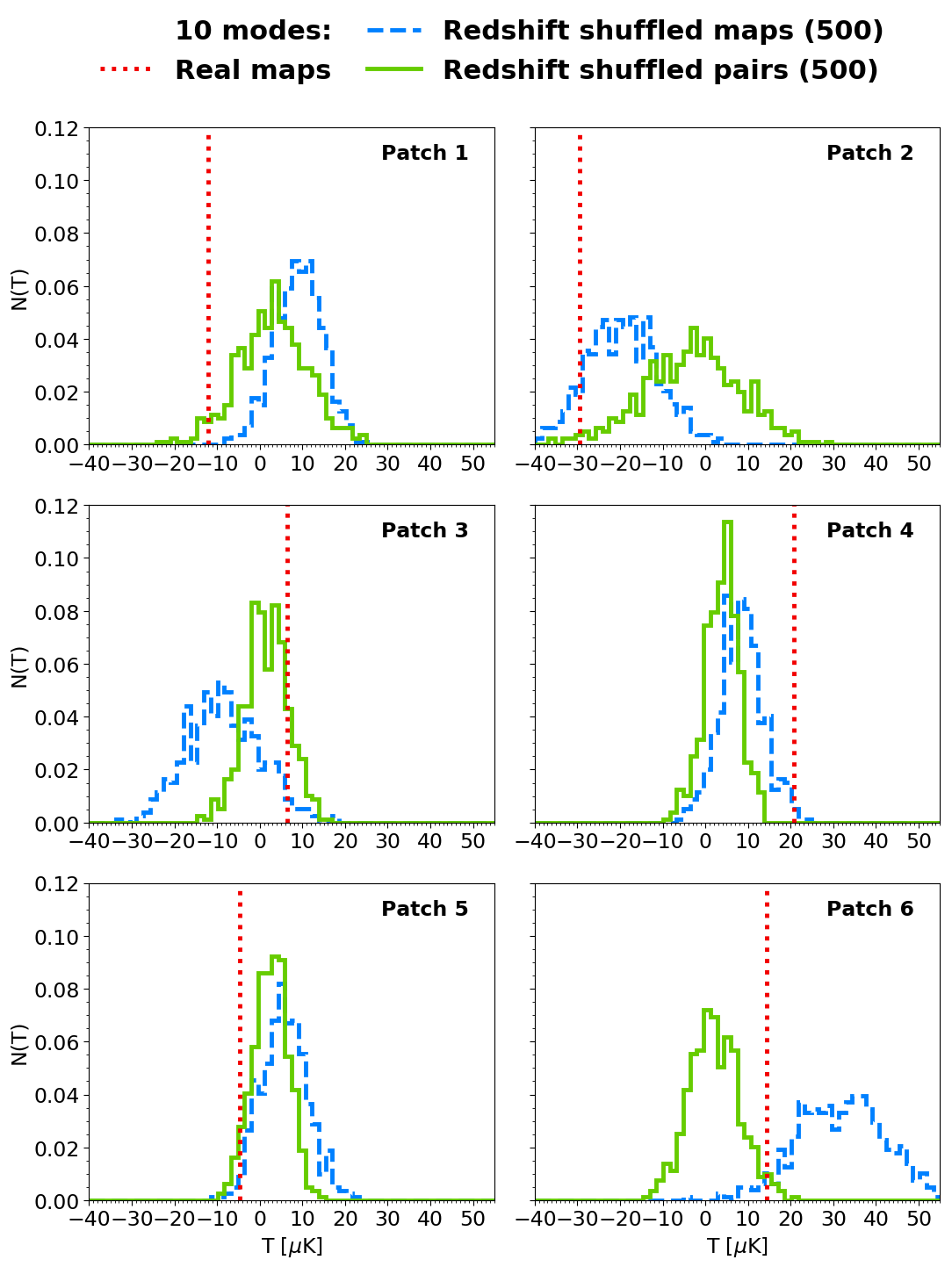}\qquad
\includegraphics[trim= 0mm 0mm 0mm 0mm, scale=0.31]{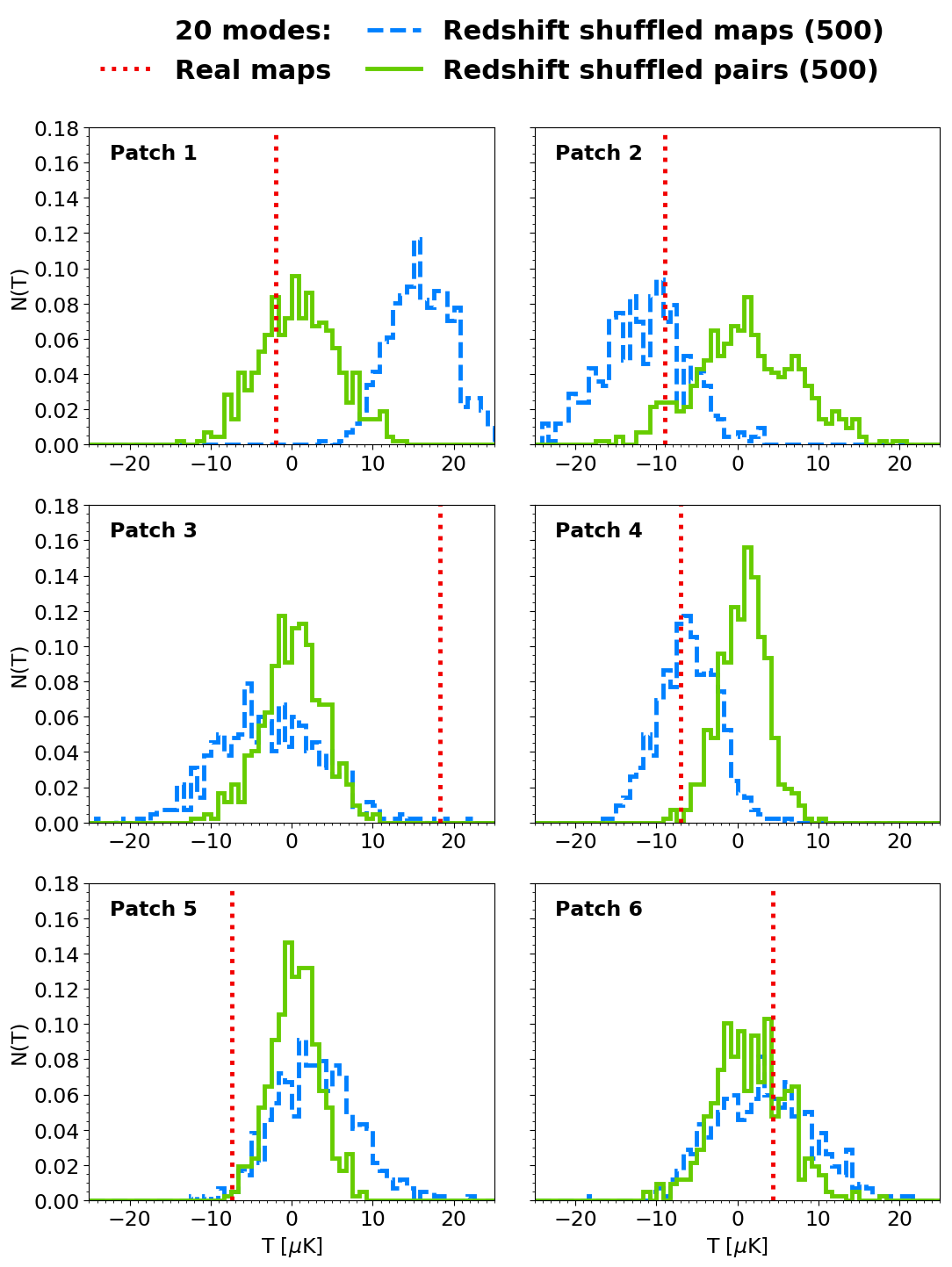}
	\caption{Significance of the filament detection shown for the six patches, for the 10-mode (\textit{left})
	and 20-mode (\textit{right}) removal cases. The vertical 
	line represents the detection on the real data, while the values obtained from the 
	500 random realisations are shown as histograms, one for each of the two methods. 
        The distributions have been normalised to unit integral.  }
\label{fig:rand_patches}	
\end{figure*}

\begin{figure*}
\includegraphics[trim= 0mm 0mm 0mm 0mm, scale=0.35]{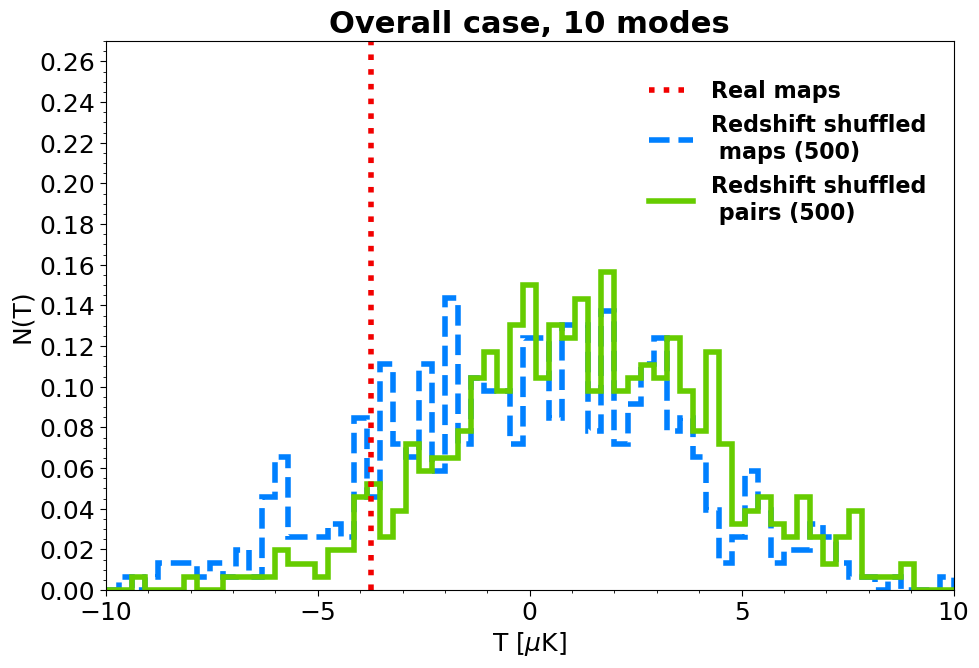}
\includegraphics[trim= 0mm 0mm 0mm 0mm, scale=0.35]{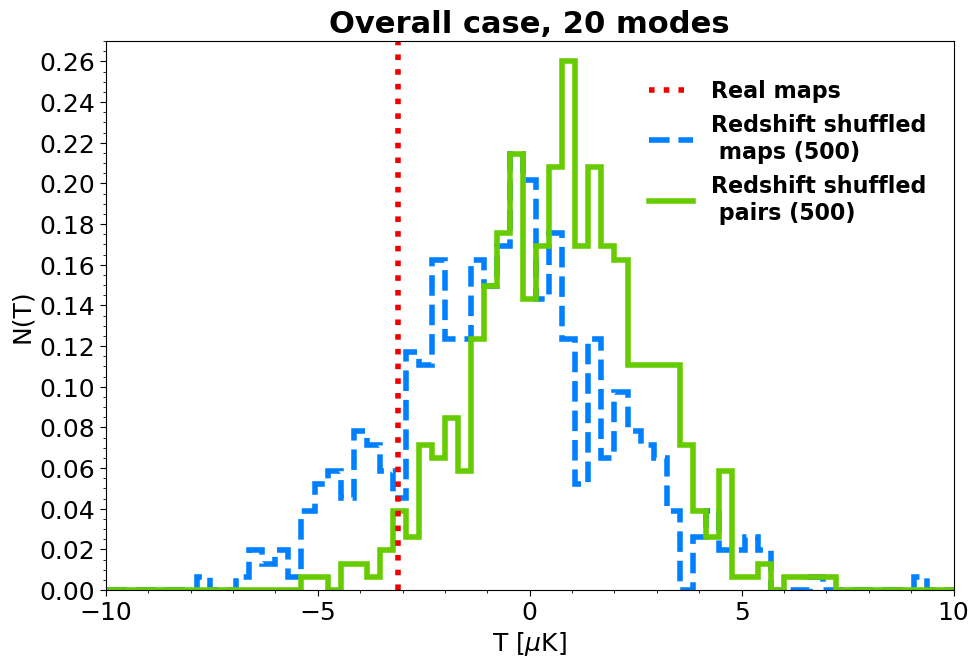}
	\caption{Same as in Fig.~\ref{fig:rand_patches}, but showing this time
	the detections and distributions on the 
	overall maps obtained by combining the contributions from the six patches.}
	\label{fig:rand_total}	
\end{figure*}

Fig.~\ref{fig:rand_realis} shows the result of the stacking for both these 
two types of randomised data sets, in the case of one particular realisation.
The result from both implementations of the bootstrap method is a set of 
500 filament values for each patch and for the overall stack, shown 
respectively in Figs.~\ref{fig:rand_patches} and~\ref{fig:rand_total} as histograms over 
the filament HI temperature.
The dispersion of these values around their mean provides an estimate for the 
uncertainty of the filament detection in the corresponding patch case. 

\section{Discussion of stacking results}
\label{sec:discussion_stack}
The initial stacks reported in Fig.~\ref{fig:stack_all} show that 
the contribution from the galactic HI emission is the dominant feature 
in the maps, and the region surrounding the galaxy pair looks quite uniform. The structures
visible in patch 1 are at the same level as in the other patches, but the lower halo-peak 
amplitude in this case enhances their contrast. The overall stacks in 
Fig.~\ref{fig:stack_total} show the improvement in the quality of the stacking 
produced by the higher available statistics, which results in a more homogeneous background. 
However, although the signal pattern looks similar for all the 
stacks, its amplitude depends on the considered patch and on the foreground removal choice, 
with the 10-mode case being systematically higher; in general, there is no regular 
trend across the patches, or a clear relation between the number of pairs 
that are stacked and the differences in the final maps.
\begin{figure}
\includegraphics[trim= 5mm 0mm 0mm 0mm, scale=0.38]{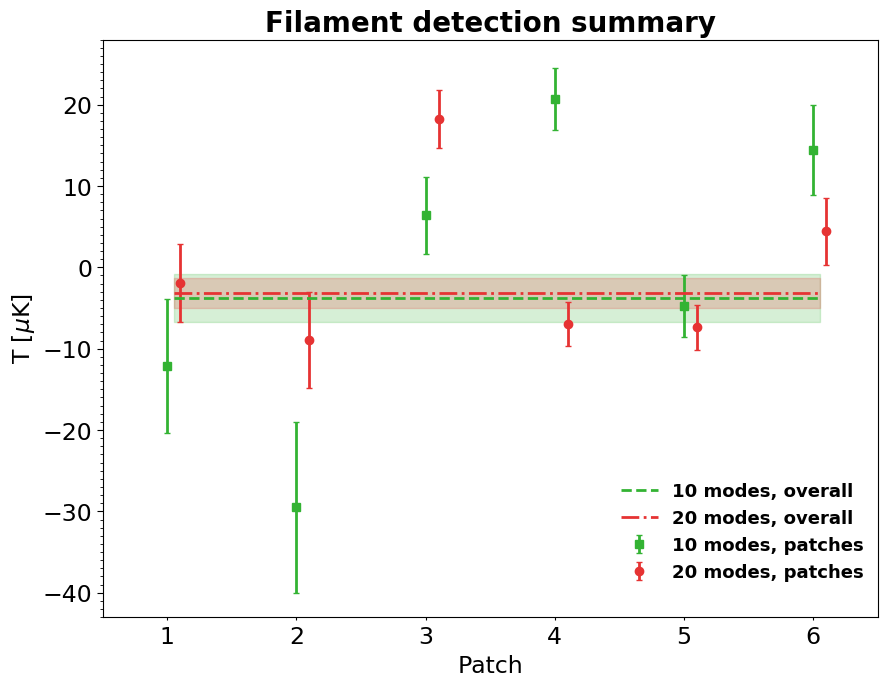}
\caption{Summary of the filament detection analysis. The filament temperatures for all the 
	patches are shown as points with their error-bars, for both the 10 and the 20-mode 
	removed cases (the two sets points are shifted 
	horizontally for displaying purposes). The estimates from the overall case that combines 
	all patches are shown as horizontal dashed lines, and their uncertainties as a shaded region. 
	Although the 20-mode removal case provides estimates closer to zero, the combination of all
	patches is compatible in the two cases. }
\label{fig:summary_fil}	
\end{figure}

At this level the presence of a filament bridging the two halo peaks is not clear, 
although some patches seem to show some overdensity in their centre, 
namely 3, 4 and 6 for the 10-mode case and 3 and 6 for the 20-mode case.
Apart from this qualitative assessment it is still not clear whether the signal excess 
is coming from a diffuse source in the map centre or whether it is just  
produced by the superposition of the two decaying halo profiles. 

The filament analysis described in Section~\ref{ssec:filament} is devoted to clarifying 
this point. The second column panels in Fig.~\ref{fig:stack_total} show examples of 
the halo profiles 
we fit over the maps; our implementation is effective in reproducing the overall shape and 
amplitude of the profiles visible in the stacks. The fact that the fit is performed 
on the outermost region of the maps, and that the two halo profiles are fitted for 
at the same time, ensures a sensible assessment of the 
emission level in the map centre coming exclusively from the profile superposition. 
Any emission excess in this region must be due to additional sources that are located 
along the line
connecting the halos, and would therefore prove the existence of a filament.
The residual maps in Fig.~\ref{fig:filament_all} again point toward different 
conclusions depending on the considered patch.  
In the presence of an actual filament these maps would clearly show a positive residual 
in their centre, with a significative contrast with respect to the rest of the map. 
This is true only for patches number 4 and 6 in the 10-mode case, and for patch number 
3 in the 20-mode case. 
All the other maps show a random distribution of positive and negative 
structures, without any emerging feature. Most importantly, the same can be said 
about the combination of all patches in the overall maps 
(Fig.~\ref{fig:stack_total}, third column). 

The one-dimensional cuts shown in Fig.~\ref{fig:profiles_all} and 
in the last column panels of 
Fig.~\ref{fig:stack_total} provide further insights for understanding the features 
visible in the residual maps. Apart from the aforementioned patches showing a clear 
positive central residual, it is clear from these plots that negative features 
in the residual maps are produced by a local overestimation of the halo profiles. 
In many cases this is due to the profile being more extended in the external part of the 
maps, and falling more abruptly towards the centre. This effect basically proves the
lack of a filament signal: although it is reasonable to assume that the halo profiles are 
circular symmetric, as we did for our fits, 
contributions from noise and possible foreground residuals 
will emerge in the final stacks as fluctuations over this symmetrical profile. When 
we subtract 
the halo signal, which is the dominant feature in the maps, these fluctuations 
will be enhanced. In the absence of a filament the final residual map signal will be at 
the level of the homogeneous background, and those fluctuations are expected to produce 
positive and negative structures, as observed in most of the stacks. In the case of a 
clear filament detection these fluctuations would still be second order compared to the 
filament signal: this is the case only for patches 4 and 6 in the 10-mode case, 
and for patch 3 in the 20-mode case. In all the other cases we can rule out the 
detection of a filament. There are other stacks with a positive 
central residual (patch 3 for the 10-mode case and patch 6 for the 20-mode case), 
but in those cases the local overdensity is comparable with the background fluctutations,
hence they cannot be considered an evidence for a filament emission. 

Regardless the appearance of the residual maps, we proceeded to the computation of the 
filament emission in the way described 
 in Section~\ref{ssec:filament} for all the stacks, and with the estimation of the 
 corresponding uncertainties as detailed in Section~\ref{ssec:randstack}.
Fig.~\ref{fig:rand_realis} shows a comparison between the two ways we implemented the 
randomisation of our data set, for one particular realisation. 
The stacking of the randomised catalogues produces
final maps with no significant features: as expected, the halo peaks are no longer visible, 
and the distribution of fluctuations is homogeneous across the full area. 
The stack with the randomised maps, although lacking the clear peak signature that 
characterises the real data maps, still shows some residual structures in correspondence 
to the nominal galaxy positions, probably due to a combination of an insufficient 
randomisation of the slices and the non-homogeneous angular and redshift distribution 
of the galaxy pairs, which tend to favour local signal patterns. 
One other observation is that the typical size of the structures visible in the shuffled 
map stacks is smaller than the ones visible in 
the shuffled catalogue stacks, which are instead similar to the residual maps shown in 
Fig.~\ref{fig:filament_all} when we have no clear filament signature;
 this effect could be produced by the breaking of frequency coherent structures when 
 shuffling the slices. The final width of the resulting distributions visible in 
 Figs.~\ref{fig:rand_patches} and~\ref{fig:rand_total} is comparable in the two cases, but 
 the residual structures in the slice shuffling method determines a shift of the mean 
 towards positive or negative values; the distributions obtained randomising the 
 pair redshifts, instead, are more symmetrically centred at zero, as expected. For 
 all these reasons we decided to discard the uncertainties estimated shuffling the maps
  and use only the ones obtained from the catalogue randomisation. 
 
Fig.~\ref{fig:summary_fil} summarises our filament detection for all individual patches 
and their combination, in the case of both 10 and 20-mode removal, together with the 
resulting uncertainties. The most evident feature emerging from this plot is the 
scatter of the values around zero, without any trend across different patches. 
Although there is some resemblance between the results obtained with different foreground 
removals, the two cases produce different estimates, with the 20-mode case being more consistently 
close to zero. 
The fact that the 10-mode removal case produces larger deviations from zero, both 
positive and negative, suggests that some foreground residuals may still affect the signal we are detecting. 
However, the random scatter across different patches indicates that these detections are probably 
dominated by noise; in these stack maps, the latter will be a combination of the initial map thermal noise and  
a number of systematic components deriving from our data processing (e.g. IPM extraction method, IPM rotation and scaling,
choice of radial bins for the profile fitting, or region chosen for measuring the filament signal).
When combining all the patches together, in any case, the 
filament temperatures estimated with different foreground removals are compatible; 
the resulting average is negative, which in the context of this stacking is non-physical.
Although some patches, as already mentioned, produce a clear filament detection, 
we cannot consider only their contribution for the overall estimate, given that we 
have no \textit{a priori} motivation for choosing some patches over others. 
Therefore, the conclusion of this analysis is that we do not detect a filament signal; this 
can be due either to the actual absence of HI in the filaments between the galaxy pairs
we are considering, or to the fact that the noise level in the final filament 
maps masks the underlying HI emission. 

In order to account for the latter case, we can quote upper limits on the HI temperature in 
the filaments; they are reported in Table~\ref{tab:uplims}, at the two-sigma level. Notice that for both 
the 10 and 20 mode cases the error for the overall stack, shown as a shaded region in Fig.~\ref{fig:summary_fil}, 
is lower than the standard error of the mean of the individual patch estimates. We decided to take a conservative 
approach and consider the mean standard error as representative of the final uncertainty associated with the filament signal in the overall stack. 
This uncertainty was used to compute the 95\% confidence level upper limits; 
the combined estimate from the 10 and 20 mode cases points toward a local filament temperature of $T_{\rm b} \lesssim 7.5\,\mu\text{K} $ .

\begin{table*}
\centering
	\caption{Upper limits obtained for the filament HI brightness temperature, the locally defined cosmological HI
	density parameter, the local overdensity of neutral baryons and the HI column density. Values are 
	reported for both the foreground removal cases, at $2\sigma$ confidence level.}
\label{tab:uplims}	
\setlength{\tabcolsep}{2em}
 \begin{tabular}{ccccc}
 \hline
 \hline
	 & $T_{\rm b}$ & $\Omega_{\rm HI}^{\rm (f)}$ & $x_{\rm HI}\,\delta_{\rm b}$ & $N_{\rm HI}$ \\ 
	 & [$\mu\text{K}$] &  $[10^{-5}]$ &  $[10^{-4}]$ & $[10^{15}\,\text{cm}^{-2}]$  \\ 
 \hline
	 10 modes & $\lesssim$ 10.3 & $\lesssim$ 7.0 & $\lesssim$ 8.2 & $\lesssim$ 4.6 \\
	 20 modes & $\lesssim$ 4.8 & $\lesssim$ 3.2 & $\lesssim$ 3.8 & $\lesssim$ 2.1 \\
 \hline
 \end{tabular}
\end{table*}


\section{Estimation of HI abundance}
\label{sec:hi}

In this section we use the estimates of the local filament excess temperature 
to infer upper limits on the HI abundance. The redshift distribution of 
the 2dFGRS catalogue, combined with the spatial separation constraints we imposed 
when defining a galaxy pair, results in a mean galactic angular separation of 
2.8 degrees, which is considerably larger than the map resolution. Since this is 
the typical scale of the signal we want to detect, the dilution effect produced by 
the Parkes beam is negligible in this case, and the antenna temperature is  
 equivalent to the local HI brightness temperature $T_{\rm b}$. For a given 
redshift $z$, the HI brightness temperature is related to the HI density parameter  
$\Omega_{\text{HI}}$ via~\citep[see for instance][]{hall13}: 
\begin{equation}
	\Omega_{\text{HI}} = 7.6\times10^{-3} \left[\dfrac{T_{\rm b}(z)}{\rm mK}\right]\,\left(\frac{h}{0.7}\right)^{-1}\left( 1+z\right)^{-2} E(z),
\end{equation}
where $E(z) = [\Omega_{\rm m}(1+z)^3 + \Omega_{\Lambda}]^{1/2}$ is the ratio between the 
Hubble parameter at redshift $z$ and its present value. Our upper limits in the 
brightness temperature can thus be converted into upper limits on the hydrogen density 
parameter, which are reported in Table~\ref{tab:uplims}. Our evaluation suggests 
$\Omega_{\rm HI}^{\rm (f)} \lesssim 5\times10^{-5}$, the upper limit being one order of magnitude lower than 
previous estimates of the HI abundance at low redshift. Observations in 21-cm of 
individual objects agree in $\Omega_{\rm HI}$ values in the range 3 to 5$\times10^{-4}$, 
obtained by fitting for the HI mass function~\citep{zwaan05, martin10, hoppmann15} and 
by stacking galaxy spectra~\citep{delhaize13, rhee13}. 
In our case, the low value of $\Omega_{\rm HI}^{\rm (f)}$ is due to the fact that it represents the local 
limit HI mass density inside a filament, but not the filament contribution to the total $\Omega_{\rm HI}$ in the Universe;
the latter could be estimated if the total number of filaments and their size were known. In this context it is more
meaningful to convert the brightness temperature into a measure of the local baryon 
overdensity $\delta_{\rm b}$; this can be constrained together with the neutral fraction of hydrogen $x_{\rm HI}$ 
using~\citep{pritchard08}:
\begin{align}
	\left(\dfrac{T_{\rm b}}{\mu\text{K}} \right) =\, &2.7\times10^4\,x_{\rm HI}\left( 1+\dfrac{4}{3}\delta_{\rm b} \right)\,\left(\dfrac{\Omega_{\rm b}h^2}{0.023}\right)   \nonumber \\ 
	&\times \left(\dfrac{0.15}{\Omega_{\rm m}h^2}\dfrac{1+z}{10}\right)^{1/2} \left(1-\dfrac{T_{\gamma}}{T_{\rm s}}\right),
\end{align}
where $T_{\gamma}$ is the CMB temperature and $T_{\rm s}$ the spin temperature. In our case, given the 
expected low density of HI in the intergalactic filaments, we are in the optically thin regime 
and $T_{\rm s} \gg T_{\gamma}$, which allows us to discard the last factor; we can also remove the unity term added to the baryon overdensity, because it
 represents the the mean HI contribution which is subtracted from our stacks.
By substituting the values for our fiducial cosmology we obtain:
\begin{equation}
	\label{eq:xdb}
	\left(\frac{x_{\rm HI}}{10^{-5}}\right) \left(\frac{\delta_{\rm b}}{10}\right) = 0.80\, \left(\frac{T_{\rm b}}{\mu\text{K}}\right),
\end{equation}
where the HI neutral fraction and baryon overdensity are normalised to the values typically expected in WHIM 
(see the discussion in Section~\ref{sec:introduction}). The substitution of our brightness temperature upper limits 
yields the estimates reported in Table~\ref{tab:uplims} for the combination $x_{\rm HI}\,\delta_{\rm b}$.
Since our overall temperature upper limit is $\sim 7.5\,\mu\text{K}$, the constraint imposed by eq.~\eqref{eq:xdb} is 
$x_{\rm HI}\,\delta_{\rm b}\lesssim 6\times10^{-4}$, which is easily satisfied by the typical WHIM values for HI neutral fraction and baryon overdensity. 
In other words, our detection upper limits provide a loose constraint for the neutral baryon overdensity in filaments.

It is also useful in this context to estimate the HI column density, defined as the 
integral of the numerical HI density along the line of sight. In the optically thin 
regime the column density is related to the brightness temperature via~\citep{meyer17}:
\begin{equation}
	\left(\frac{N_{\rm HI}}{\text{cm}^{-2}}\right) = 1.82\times10^{15}\, \int \,\text{d}\left(\dfrac{\text{v}}{\text{km}\,\text{s}^{-1}}\right)\,\left[\frac{T_{\rm b}(\text{v})}{\text{mK}}\right],
\end{equation}
where $T_{\rm}(\text{v})$ is the observed brightness at radial velocity $\text{v}=cz$, 
and the integral extends over the line profile. 
By using $1+z=\nu_{21}/\nu$, with $\nu_{21}$ the 21-cm line rest frequency and $\nu$ the 
observed frequency, the relation above can be recast as:
\begin{equation}
	\left( \frac{N_{\rm HI}}{\text{cm}^{-2}}\right) = 4.48\times10^{14}\, \int \,\text{d}\left(\dfrac{\nu}{\text{MHz}}\right)\,\left[\frac{T_{\rm b}(\nu)}{\mu\text{K}}\right].
\end{equation}
The integral can be approximated by the product of the integrand evaluated in the mean 
frequency of the maps, $\nu = 1315.5\,\text{MHz}$, and the frequency channel width, 
$\Delta\nu=1\,\text{MHz}$; by inserting the measured upper limits in the HI brightness 
temperature, we obtain the estimates reported in Table~\ref{tab:uplims}, which 
average around $3.4\times10^{15}\,\text{cm}^{-2}$. This value is significantly lower than 
the HI abundance found in halos; damped Ly $\alpha$ systems, for instance, generally have 
column densities $N_{\rm HI} > 10^{22}\,\text{cm}^{-2}$~\citep{bird17}. Our estimate, however, is still 
about one order of magnitude higher than the HI content we would expect to observe in filaments, given that HI detections 
in WHIM have been performed at lower column densities~\citep{richter06, nicastro13}. Again, although
not representing a tight upper limit, our constraint is consistent with previous determination of the filament HI abundance.

The column density can also be evaluated using the estimates we obtained on the
HI density parameter, provided the shape of the filaments is known. 
Assuming that the HI is uniformly distributed inside the filament, we can write:
\begin{equation}
	N_{\rm HI} = \frac{\Omega_{\rm HI}^{\rm (f)} \rho_{c}}{m_{\rm HI}} (1+z)^3 \,\Delta s,
\end{equation}
where $m_{\rm HI}$ is the hydrogen atomic mass, $\rho_{c}$ is the 
Universe critical density and $\Delta s$ is filament extension 
along the line of sight. By substitution of the numerical values, and in 
astrophysical convenient units, the latter relation reads:
\begin{equation}
	\left(\frac{N_{\rm HI}}{10^{15}\,\text{cm}^{-2}}\right)= 0.21\left(\frac{\Omega_{\rm HI}^{\rm (f)}}{10^{-5}}\right) \left(\frac{\Delta s}{\text{Mpc}}\right).
\end{equation}
In order to match the estimates in $\Omega_{\rm HI}^{\rm (f)}$ and $N_{\rm HI}$ 
reported in Table~\ref{tab:uplims} the filament thickness is constrained to $\sim 3\,\text{Mpc}$.
There is no universal definition for the typical filament radial extension;  
galaxies are usually considered filament members up to a radial distance of 
$0.5\,h^{-1}\text{Mpc}$~\citep{tempel07} or $1\,h^{-1}\text{Mpc}$~\citep{kooistra17}
from the filament spine. The latter is consistent with our constraint if we take 
$\Delta s$ to represent the diametral filament size. This rough estimate serves as 
a consistency check of our estimated HI abundance. 

\subsection{Comparison with results from numerical simulations}

To date there is no direct detection of neutral hydrogen in filaments from 21-cm observations;
as mentioned in Section~\ref{sec:introduction}, most of the observational efforts have been devoted to 
the study of the ionised gas, while the detection of the neutral component relied on the 
observations of BLAs in the UV range. The detection of the HI 21-cm emission at the column density levels quoted in Table~\ref{tab:uplims} 
is currently not feasible, and is the goal for the next generation of HI surveys.
In this context, the use of numerical simulations becomes of paramount importance in order to provide 
hints on the conditions of the neutral gas in filaments; this information is also useful in assessing the best 
observational strategy to detect the faint filamentary HI signal. As far as the present work is concerned, 
results from simulations provide an interesting benchmark to compare our results. 

We have already mentioned in the introduction that it is actually the study of cosmological hydrodynamical simulations that 
initially established the WHIM component as the most likely reservoir of missing baryons at low redshift. 
The first results from the seminal works by~\citet{cen99} and~\citet{dave99} have been updated during the last two decades, 
following the improvements in the implementation of the baryon physics in the simulations, leading to the latest studies  
by~\citet{haider16},~\citet{cui19} and~\citet{martizzi19}. The latter work, in particular, was based on state-of-the-art 
numerical simulations that include a detailed characterisation of the baryonic physics involved in galaxy formation, 
evolution and feedback; the study explored the distribution of distinct baryonic phases, defined by 
cuts in the density-temperature phase diagram, across different structures of the cosmic web, namely nodes, filaments, 
sheets and voids. Qualitatively, the study reiterated the conclusion that the WHIM gas is mostly found at low redshift 
in filaments and nodes: at $z=0$ filaments account for the majority of the baryonic mass, where it is found 
as a combination of the WHIM and intergalactic medium (IGM) phases. The IGM in their work is identified as the gas component with 
density comparable to the WHIM, but with lower temperature ($T \lesssim 10^5\,\text{K}$). They also found that the IGM 
is rather ubiquitous across different cosmic structures, whereas the WHIM component is concentrated
in the proximity of LSS filaments and nodes. The WHIM is therefore an effective tracer of the LSS filaments. 
Although the IGM component carries 
both ionised and neutral hydrogen, their work did not explicitly quote the final fraction of HI in filaments.

Studies with focus on the neutral baryonic component have also been tackled in the literature. 
In particular,~\citet{popping09} showed how cosmological hydrodynamical simulations can conveniently complement  
real data below the current observational sensitivities. Starting from the output 
of a cosmological hydrodynamical simulation, they derived statistical properties for the hydrogen distribution which agree with 
observational constraints in the column density range for which data are available; this enables to study the simulated 
predictions below the observable limits. More precisely, they reconstructed the three-dimensional distribution of 
the total, neutral atomic and molecular hydrogen components, and extracted two dimensional maps for 
the corresponding column densities. The total hydrogen component clearly traces the typical cosmic web configuration 
and yields very high column densities in correspondence to the nodes, with peak values of 
$N_{\rm H}\sim10^{21}\,\text{cm}^{-2}$, while the filaments are associated with values typically one order of magnitude 
lower. The $\text{H}_2$ column density map loses all the diffuse structures and exclusively retains the LSS nodes, 
which are the only locations where density and pressure are high enough to allow the formation of the molecular phase. This
molecular gas generally hosts star formation, and the nodes correspond to massive galaxies and groups. The HI map also 
shows very high column densities in correspondence with the LSS nodes, but the signal is considerably lower for 
the filaments; nonetheless, the neutral atomic hydrogen still traces the large scale distribution of baryons, 
and in filaments the typical column density is $N_{\rm HI}\sim10^{16}\,\text{cm}^{-2}$. This value is almost one order of 
magnitude higher than our upper limits from Table~\ref{tab:uplims}; we have to stress, however, that such an estimate 
is quoted in the aforementioned reference as a general value, but there is no clear explanation on how it has been obtained. From their map of the 
HI column density, filaments can clearly be identified at lower values of $N_{\rm HI}$, so their quote can be considered  
representative of the brightest filaments. More interestingly, they described a correlation between the HI column density 
and the expected neutral fraction of hydrogen $x_{\rm HI}$, with the gas being almost fully neutral at $N_{\rm HI}\gtrsim 10^{20}\,\text{cm}^{-2}$, 
but showing a steep increase in the ionised fraction as the column density decreases. 
When placing the values extracted from the simulation on a $x_{\rm HI}$--$N_{\rm HI}$ plane, the points scatter around a well defined 
curve; it is then meaningful to ask what is the most likely occurrence for the combination of neutral fraction and column 
density, as the region of the plot with the highest density of points. The conclusion is that  
 $N_{\rm HI}\sim10^{14}\,\text{cm}^{-2}$ and $x_{\rm HI}\sim 10^{-5}$  are the most commonly occurring conditions. At a column 
density of $N_{\rm HI}\sim10^{15}\,\text{cm}^{-2}$, which is similar to our upper limit, their relation suggests a value for the neutral fraction 
 just above $10^{-5}$, and, qualitatively, the probability of occurrence of this combined condition is still comparable to the maximum.
Our upper limit on the combination $x_{\rm HI}\delta_{\rm b}$ is then consistent with this result if the filament overdensity 
is $\delta_{\rm b}\lesssim 60$. Since the local filament overdensities are typically in the range 10 to 100, our findings agree with the simulation results.

There have been subsequent studies focused on the prediction of properties of the neutral baryonic component based on 
numerical simulations~\citep{duffy12,cunnama14}, although most of them do not target specifically the study of HI in 
filaments. 
The detectability of filamentary neutral gas using 21-cm observations has been investigated by~\citet{takeuchi14}. In that 
work the evolution of the abundance of different hydrogen ionisation states and the gas temperature in the IGM 
are at first computed analytically; the results enable the prediction of the typical HI brightness temperature and column density expected in filaments. 
Assuming a baryon overdensity of $\delta_{\rm b}=100$, the resulting column density for a filament which extends over $1\,h^{-1}\text{Mpc}$ 
along the line of sight is quoted to be $N_{\rm HI}\simeq 10^{15}-10^{16}\,\text{cm}^{-2}$, and assuming a proper LoS velocity
of $\Delta \text{v}=300\,\text{km}\,\text{s}^{-1}$, the corresponding observed brightness temperature lies in the range $T_{\rm b}\simeq 10^{-6}-10^{-5}\,\text{K}$.
Given the mean redshift and the bandwidth of the Parkes data we employed, in our case the mean LoS velocity is 
$\Delta \text{v} \simeq 245\,\text{km}\,\text{s}^{-1}$, implying a final brightness temperature a factor 1.2 lower; the results is 
still consistent with their estimates, and since their computed brightness temperature is proportional to the baryon overdensity and 
the filament LoS extension, we can easily accommodate a lower overdensity of $\delta_{\rm b}\sim 50$, consistent with 
the previous discussion about the neutral fraction, and a larger filament extension of $\sim 2.8\,\text{Mpc}$, comparable with our 
internal consistency check for the column density computed using the HI density parameter in filaments $\Omega_{\rm HI}^{\rm (f)}$. 
The study in~\citet{takeuchi14} continues 
by comparing these estimates with the direct measurement of the filament properties on N-body simulations; they extracted the 
filament density contrast and LoS velocity from snapshots of the simulation at different redshits, and used them to estimate the 
brightness temperature; at $z=0$ they found that actually $\delta_{\rm b} \lesssim 100$, which yields a lower brightness temperature, 
at the $\mu\text{K}$ level. Again, this is in agreement with our upper limits. 

The work by~\citet{horii17} reprised this type of analysis, employing cosmological hydrodynamical simulations instead 
of N-body simulations; they stressed that the use of the latter would likely lead to an overestimation of the neutral fraction 
of gas in the filaments, due to the lack of feedback effects from galaxies. In any case, their results agree qualitatively with 
the previous ones, and show that HI is effective in tracing the filaments in the cosmic web; in the corresponding brightness temperature 
map at $z=0.5$, the signal shows peaks at the level of $10^{-4}\,\text{K}$ in the nodes, and lower values of $10^{-5}-10^{-6}\,\text{K}$ in the filaments.
The comparison with the same map at $z=1$ shows that we can only expect a very mild decrease of these estimates to $z=0$, so that once 
again we find agreement with our upper limits. The work continues with the discussion on the possible contribution to the 
observed signal by filament galaxies, and provides an estimate for the brightness temperature coming from the diffuse WHIM component alone, 
resulting in even lower values of $\sim 10^{-8}\,\text{K}$ obtained by masking higher overdensity regions in the filaments. The detection 
of such a faint signal is clearly not feasible with any of the upcoming HI surveys, and will require the latest phases of the Square Kilometer 
Array~\citep{braun15}; they also 
stated, however, that this result is affected by uncertainties regarding the actual contribution from filament galaxies. 

Clearly, the detailed predictions from numerical simulations depend on the modelling of the physics 
underlying the formation and evolution of structures, and are also affected by the choice of the box size and resolution. 
Despite these drawbacks, they still represent a convenient gauge to test results like the ones presented in this study, 
in the absence of direct observational constraints. 
As far as our work is concerned, we find a general agreement between our upper limits and the simulated properties of HI at very low redshift.
The continuous improvement in the modelling, resolution and size of numerical simulations can clearly provide a deeper understanding of the HI state in WHIM, although 
the ultimate confirmation can only be achieved with the next generation of 21-cm surveys.


\section{Conclusions}
\label{sec:conclusions}

This work attempts at finding HI emission from filaments connecting halos in the 
large-scale structure of matter in the Universe. In the framework of the missing 
baryon problem, the results of this analysis aim at constraining the amount of neutral 
hydrogen that can be found in the diffuse WHIM component. Previous work provided 
estimates based on BLA detection, but were limited to the study of a relatively small 
sample. 
This work addresses the analysis of the HI filament content by  
joining the contribution of a large number of objects distributed over an 
extended sky area ($\sim 1,300$ square degrees). It represents the analogous 
of the search for baryons that~\citetalias{degraaff19} and~\citetalias{tanimura19} 
conducted on the Compton parameter map, but using for the first time the 21-cm 
signal of the HI component. It takes advantage of the capabilities of the low resolution  
HI intensity mapping technique of covering large areas on the sky, and resorts to a 
stacking procedure to enhance the signal that we are looking for. 

We located the filament positions by selecting suitable pairs of 
2dFGRS galaxies which define their endpoints, and measured the 21-cm signal on a set of Parkes maps 
produced by observations that spanned the 2dF survey area. For each individual pair we 
built a local two-dimensional 21-cm map by projecting a plane joining the two galaxies along 
the line of sight, and by rotating and scaling the local field of view in order to bring all the 
pairs to a common frame. We then co-added together these maps in order to enhance the filament 
contribution, and removed the HI signal proceeding from the pair galaxies by jointly fitting 
a pair of circular symmetric one-halo profiles on the final stacks. By joining the contribution of the 
six Parkes data patches, we stacked a total of 274,712 galaxy pairs with a mean redshift 
$z = 0.079$. The final stacks clearly showed the contribution from the galactic HI emission, 
but once the fitted halo profiles were removed, the resulting residuals did not provide a clear 
evidence for a filament emission. Nonetheless, we estimated the local filament HI brightness 
temperature as the mean value of the residual maps in their central region, and assigned to each 
estimate an uncertainty determined as the variance of the residual signal from a set of 500 
random realisation of the same analysis, obtained by stacking redhift shuffled versions of the 
galaxy pair catalogue.

The final values for the filament brightness temperature show a random dependence on the 
considered sky patch, with the results from the 20 foreground modes removed maps being 
more consistently close to zero than the 10-mode case. This suggests that a possible 
filament signal is at the same level of, or lower than, the noise fluctutations 
in the final maps, to which residual foregrounds may contribute to some extent; this conclusion 
is corroborated by the fact that many estimates are 
negative. The overall stack obtained combining all patches shows consistency across the 
two foreground removal cases, within the error bars generated with the bootstrap 
method. Since these errors are 
 lower than the standard deviation of the mean computed from the individual patch detections, 
we considered the latter to be a more sensible estimate for the measured filament residual uncertainty. 
This allows us to set an overall upper limit for the filament HI 
brightness temperature at $T_{\rm b}\sim 7.5\,\mu\text{K}$, which is converted into a local
filament HI cosmological density parameter as $\Omega_{\rm HI}^{\rm (f)}\sim 5\times10^{-5}$ and column density as 
$N_{\rm HI}\sim 3.4\times 10^{15}\,\text{cm}^{-2}$. The value of $\Omega_{\rm HI}^{\rm (f)}$ is
low compared to previous low-redshift estimates because it represents the 
local HI density parameter inside the filaments, but not the filament 
contribution to the global $\Omega_{\rm HI}$ in the Universe. 
The constraint on $N_{\rm HI}$ is one order of magnitude higher than the typical 
	 HI column density expected from observations of BLA absorbers in the UV range; although not representing 
	 a tight upper limit, it is consistent with previous HI detections in WHIM.
The brightness temperature also allowed us 
to derive a joint constraint on the filament baryon overdensity and neutral fraction; again,  
the resulting upper limit easily accommodates the expected WHIM physical conditions. 
These results suggest that, if any actual HI component is present in the filaments 
we considered, its 21-cm emission lies at a level that cannot be detected  
with the set of maps employed in this work. Finally, we compared our estimates with 
results obtained from cosmological hydrodynamical simulations. Although the predicted properties of the low-redshift 
HI diffuse gas vary between different studies, simulations generally agree in showing how neutral hydrogen can 
effectively trace the cosmic web at $z=0$, and provide a useful benchmark in the absence of observational constraints; 
we found our upper limits to be in agreement with the simulated HI properties at $z\simeq 0$.

Notice that this paper analysis can easily be adapted to a different data set, 
provided extended 21-cm maps have a significant overlap with a chosen galaxy catalog. 
In particular, it would be interesting to conduct the same study at higher redshift, 
where the fraction of neutral gas is expected to be larger. Observations with the GBT 
telescope centred at $z\sim 0.8$ have been used in~\citet{chang10} 
and~\citet{masui13}, in cross-correlation with optical surveys; however, the sky area 
explored in those works is considerably smaller than the one we employed. 
At very low redshift, future work can be addressed to tighten the upper limits 
on the HI filament content.
The ALFALFA survey~\citep{haynes18} conducted with the Arecibo telescope provides 
HI detections at $z < 0.06$; however, the legacy data consist of individual object 
catalogues, and not in continuous maps. 

In conclusion, the search for HI in filaments using HI intensity mapping and stacking techniques
requires extended maps in the 21-cm emission. 
Being able to analyse a large sky area is crucial in increasing the statistics 
of the galaxy pairs, and in enhancing a possible filament signal; besides, 
 being able to explore different sky regions would lead to more sensible conclusions, 
 mitigating the effects of local sample variance.
Surveys conducted using next-generation facilities like the
Square Kilometer Array~\citep{braun15} and the FAST telescope~\citep{li16} are expected to provide 
improved maps over a larger sky area thanks to the higher sensitivity and survey speed. 
A proper foreground removal clearly remains the critical point for delivering final 
maps with sufficient sensitivity to enable a detection of the filament HI signal.

\section*{Acknowledgements}

DT acknowledges financial post-doctoral support from the South African Claude Leon Foundation. YZM would like to acknowledge the supports from National Research Foundation with grant no. 105925 and 110984, and from the National Science Foundation of China, no. 11828301. The Parkes radio telescope is part of the Australia Telescope National Facility which is funded by the Australian Government for operation as a National Facility managed by CSIRO.
The authors would like to thank Dr. Christopher Anderson for granting access to the processed Parkes 21-cm maps, and for several valuable suggestions about the analysis presented in this work.




\bibliographystyle{mnras}
\bibliography{bibliography} 



\bsp	
\label{lastpage}
\end{document}